\documentclass[11pt,a4paper]{article}
\usepackage[T1]{fontenc}
\usepackage[utf8]{inputenc}
\usepackage{lmodern}
\usepackage[margin=23mm]{geometry}
\usepackage{amsmath,amssymb,bm,mathtools}
\usepackage{graphicx,booktabs,array,tabularx}
\usepackage{microtype}
\usepackage[numbers,sort&compress]{natbib}
\usepackage{caption,subcaption}
\usepackage[hidelinks]{hyperref}
\usepackage{placeins}
\hypersetup{pdftitle={Thermosolutal Marangoni instabilities in a weakly viscoelastic falling film with soluble surfactant},pdfauthor={Sanghasri Mukhopadhyay, Severine Millet, Bastien Di Pierro, Asim Mukhopadhyay}}
\graphicspath{{figures/}}
\newcommand{\dd}{\mathrm{d}}
\title{Thermosolutal Marangoni instabilities in a weakly viscoelastic falling film with soluble surfactant}
\author{\parbox{0.96\textwidth}{\centering
Sanghasri Mukhopadhyay\textsuperscript{1,*}, S\'everine Millet\textsuperscript{2},\\[3pt]
Bastien Di Pierro\textsuperscript{2}, and Asim Mukhopadhyay\textsuperscript{3}\\[7pt]
\small\textsuperscript{1}Centre for Applied Sciences (CAS), International Institute of Information Technology, Bengaluru 560100, India\\[3pt]
\textsuperscript{2}Universit\'e Claude Bernard Lyon 1, LMFA, UMR 5509, CNRS, Ecole Centrale de Lyon, INSA Lyon, 69622 Villeurbanne, France\\[3pt]
\textsuperscript{3}Department of Mathematics, Vivekananda Mahavidyalaya, Burdwan 713103, West Bengal, India\\[5pt]
\textsuperscript{*}\href{mailto:sanghasri.m@iiitb.ac.in}{sanghasri.m@iiitb.ac.in}}}
\date{26 October 2026}
\begin{document}
\maketitle
\begin{abstract}
We examine the linear stability and nonlinear evolution of a heated falling film containing a soluble surfactant and a weakly viscoelastic liquid. The stress response is described by the short-memory Walters model, while surface tension depends on temperature and surface concentration. An Orr--Sommerfeld analysis and its long-wave limit determine how elastic memory and the two Marangoni stresses enter the instability threshold. A five-field weighted-residual model is then derived to describe finite-amplitude disturbances. The bulk and surface transport equations are written with compatible fluxes, which conserve their combined projected surfactant inventory in a periodic domain. The numerical examples show that solutal Marangoni stress suppresses disturbance growth and reduces the amplitude of the nonlinear waves. Thermal Marangoni stress increases the maximum growth rate, although its effect on the width of the unstable wavenumber band is different. Elastic memory changes both the onset and the finite-wavelength response. The nonlinear profiles develop a main hump and capillary ripples, accompanied by variations in bulk concentration, surface concentration and interfacial temperature. These results describe the competition between thermocapillary forcing, surfactant redistribution and elastic stress within the retained long-wave model. The larger elastic parameters are interpreted as formal predictions of the truncated constitutive equations, rather than as a controlled extension to strongly elastic flow.
\end{abstract}
\noindent\textbf{Keywords:} falling films; soluble surfactants; thermocapillary instability; weak viscoelasticity; weighted-residual modelling; conservation.
\section{Introduction}\label{sec:introduction}

Gravity-driven liquid films are used in coating, absorption and heat-transfer processes in which the free surface provides a large area for exchange with the surroundings. Their usefulness also depends on the waves that develop when the uniform flow becomes unstable. A disturbance changes the local thickness, velocity and temperature fields, while surface-tension gradients introduce further stresses along the interface. Long-wave and depth-averaged descriptions provide a practical means of following these coupled processes beyond the initial growth of disturbances \cite{oronLongscaleEvolutionThin1997,changComplexWaveDynamics2002,kalliadasisFallingLiquidFilms2012}. Their accuracy depends on the approximation of the velocity and transported fields, as well as on the balance laws retained during reduction.

Heating introduces a second source of interfacial motion through the dependence of surface tension on temperature. Goussis and Kelly \cite{goussisSurfaceWaveThermocapillary1991} analysed the interaction between surface-wave and thermocapillary instabilities in a falling film. Kalliadasis \textit{et al.} \cite{kalliadasisThermocapillaryInstability2003} studied wave formation on a uniformly heated incline using an integral-boundary-layer model. Subsequent weighted-residual descriptions introduced independent velocity and thermal fields and compared their stability and wave predictions with more complete formulations \cite{ruyerquilThermocapillaryLongWaves2005,scheidThermocapillaryLongWaves2005,trevelyanHeatedFallingFilms2007}. These studies show why the interfacial temperature should generally be determined dynamically rather than prescribed as a function of the local thickness.

A soluble surfactant adds transport in both the bulk liquid and the interface. Adsorption and desorption exchange material between them, and the resulting surface-concentration gradients produce solutocapillary stresses. Karapetsas and Bontozoglou \cite{karapetsasPrimaryInstabilityFalling2013} examined the primary instability of falling films containing soluble surfactants. Pascal \textit{et al.} \cite{pascalStabilityInclinedFlow2019} developed linear and nonlinear descriptions with concentration-dependent density. For a heated Newtonian film, D'Alessio \textit{et al.} \cite{dalessioMarangoniInstabilitiesAssociated2020} combined thermal and solutal Marangoni effects in an Orr--Sommerfeld analysis and a weighted-residual model. D'Alessio and Pascal \cite{dalessioLongwaveInstabilityThin2021} subsequently studied a related long-wave formulation. These Newtonian problems provide useful limits for the present analysis.

Weak elastic memory changes the stress response without requiring a model intended for strongly elastic liquids. We use the short-memory approximation usually termed Walters' liquid B$''$ \cite{waltersMOTIONELASTICOVISCOUSLIQUID1960,beardElasticoviscousBoundarylayerFlows1964}. Applications to falling films include the studies of Andersson and Dahl \cite{anderssonGravitydrivenFlowViscoelastic1999}, Uma and Usha \cite{umaDynamicsThinViscoelastic2006}, and Mukhopadhyay and Mukhopadhyay \cite{mukhopadhyayWavesInstabilitiesViscoelastic2020}. Amatousse \textit{et al.} \cite{amatousseTravelingWavesFalling2012} derived a two-equation weighted-residual model for the clean isothermal case, while Fu \textit{et al.} \cite{fuInstabilityWeaklyViscoelastic2018} considered a heated weakly viscoelastic film using weighted-residual and Benney-type descriptions. These studies provide the clean-film and thermal benchmarks for examining the coupling to soluble-surfactant transport and a conservative reduced exchange closure.

Thermosolutal instabilities also occur in polymeric binary layers. Sarma and Mondal \cite{sarmaThermosolutalMarangoni2021}, for example, considered heating from the free surface and concentration variations associated with the Soret effect. Their initially quiescent Maxwell liquid differs from the gravity-driven film considered here. In the present problem, concentration gradients arise from the transport of a soluble surfactant and its exchange with the interface; no Soret term is included. We assume constant bulk density and viscosity, a uniformly heated impermeable wall, and heat loss to a passive gas through Newton's law of cooling.

The reduction of the transport equations requires particular care. Equal and opposite adsorption--desorption sources are necessary, but their cancellation alone does not guarantee conservation of the combined inventory. Earlier work corrected a nonlinear surface Marangoni contribution \cite{mukhopadhyayConservationConsistentBoundary2026,mukhopadhyaySolutocapillaryInstabilitySlipping2026}. A separate audit identified a remaining bulk-diffusion residual associated with depth integration at a sloping boundary \cite{mukhopadhyayBulkSurfaceAudit2026}. We retain both corrections and write the coupled reduced balances in flux form. On a periodic domain they conserve the projected inventory $\int_0^L(B+\Gamma)\,\mathrm{d}x$, where $B$ is the depth-integrated bulk concentration. This statement concerns the stated reduced model; compatibility with the full curved-interface geometry and the accuracy of the concentration profile remain separate questions.

We first derive the linear eigenvalue problem and its long-wave limit to identify the separate contributions to the onset of instability. The numerical examples then examine neutral curves and growth rates with thermal and solutal Marangoni stresses introduced separately and together. A five-field weighted-residual model provides the nonlinear description. The computed profiles show how the free surface, flow rate, surfactant distribution and temperature develop as the initial disturbance grows. Particular attention is given to the distinction between a saturated wave amplitude and a permanent travelling wave, and to the range in which the short-memory approximation can be interpreted quantitatively.

Section~\ref{sec:formulation} states the governing equations, scaling and base state. Section~\ref{sec:linear} develops the linear analysis, and Sec.~\ref{sec:nonlinear} gives the reduced nonlinear model. The subsequent sections describe the numerical procedures, examine the computed results and state the limitations of the short-memory and depth-averaged approximations.
\section{Mathematical formulation}\label{sec:formulation}

In this section we formulate the governing equations for a heated, soluble-surfactant-laden falling film of a weakly viscoelastic liquid. Dimensional variables are used in Secs.~\ref{subsec:physical_config}-\ref{subsec:surfactant_surface_law}; the non-dimensional variables and parameters are then introduced in Sec.~\ref{subsec:nondimensionalization}.

\subsection{Physical configuration}\label{subsec:physical_config}

\begin{figure}[t]
    \centering
    \includegraphics[width=0.90\linewidth]{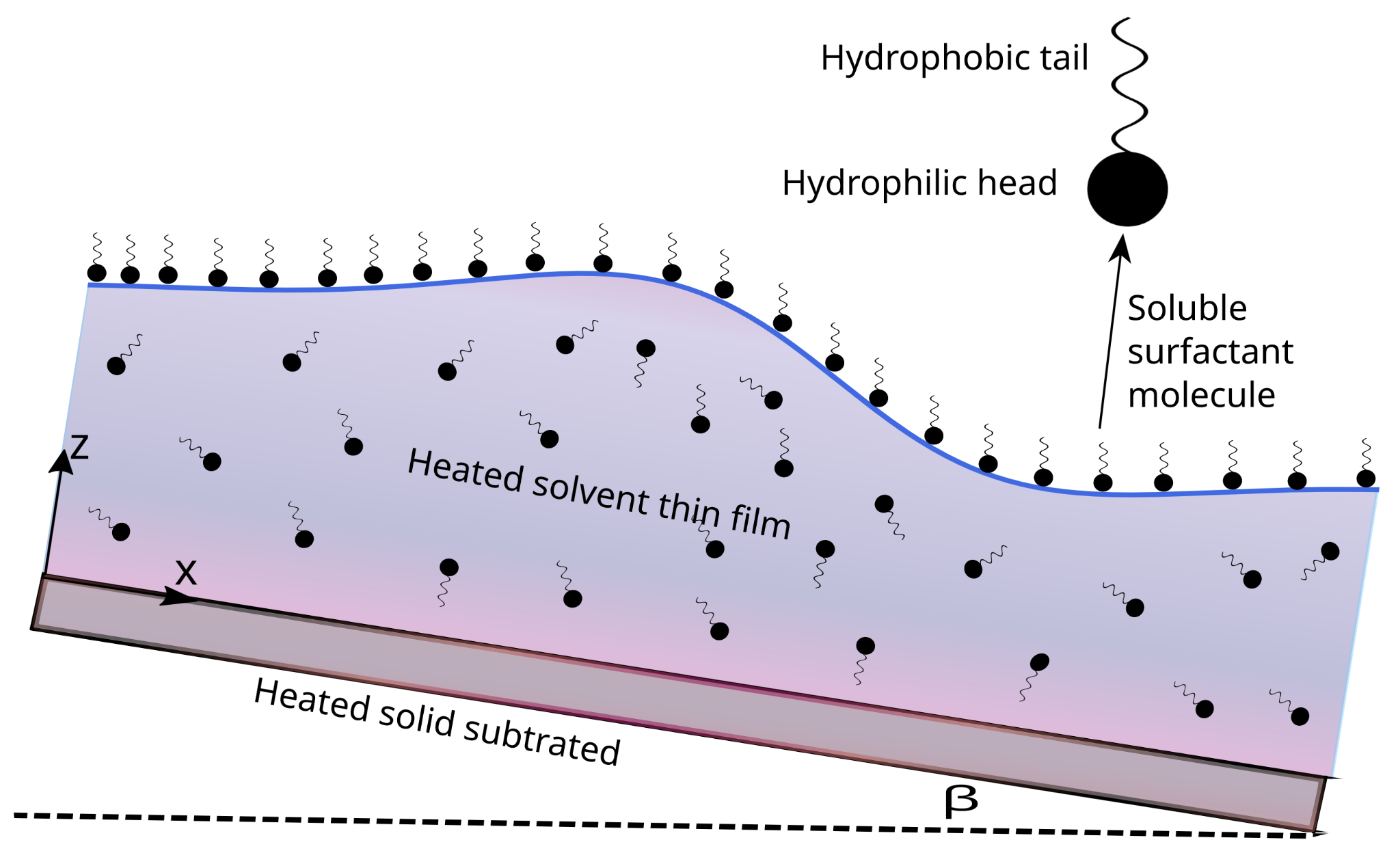}
    \caption{Schematic of a heated inclined falling film laden with soluble surfactant. The free surface is located at $z=h(x,t)$, and the plane is inclined at an angle $\beta$ to the horizontal.}
    \label{fig:diagram}
\end{figure}

We consider the two-dimensional flow of an incompressible weakly viscoelastic liquid film down a uniformly heated plane inclined at an angle $\beta$ to the horizontal, as shown in Fig.~\ref{fig:diagram}. The Cartesian coordinate $x$ is measured along the plane and $z$ is measured normal to it. The liquid-air interface is described by $z=h(x,t)$,  where  $  
h$ is the local film thickness. The velocity field is denoted by $ \mathbf{V}=(u,w),$ where $u$ and $w$ are the streamwise and cross-streamwise velocity components, respectively. The liquid has density $\rho$, dynamic viscosity $\mu$, thermal conductivity $k_T$, and specific heat $c_p$. The acceleration due to gravity is $\mathbf{g}=(g\sin\beta,-g\cos\beta).$ The wall is maintained at temperature $T_w$, while the surrounding passive gas is at temperature $T_\infty$. A soluble surfactant is present in the liquid. Its bulk concentration is denoted by $C(x,z,t)$, and its surface concentration at the liquid-air interface is denoted by $\Gamma(x,t)$.

\subsection{\texorpdfstring{Constitutive model for Walters' liquid B$''$}{Constitutive model for Walters' liquid B double prime}}\label{subsec:walters_model}

The liquid is modelled as Walters' liquid B$''$, which is a weakly elastic model appropriate for liquids with short or rapidly fading memory. This model may be regarded as a first-order correction to Newtonian behaviour and contains only one additional material parameter. It has been used previously for weakly viscoelastic boundary-layer and film-flow problems \cite{waltersMOTIONELASTICOVISCOUSLIQUID1960,beardElasticoviscousBoundarylayerFlows1964,anderssonGravitydrivenFlowViscoelastic1999,umaDynamicsThinViscoelastic2006,mukhopadhyayWavesInstabilitiesViscoelastic2020}. The total stress tensor is written as
\begin{equation}\label{eq:constitutive_dimensional}
    \boldsymbol{\tau}=-p\mathbf{I}+2\mu\mathbf{e}-2\Upsilon_0\frac{\delta\mathbf{e}}{\delta t},
\end{equation}
where $p$ is the pressure, $\mathbf{I}$ is the identity tensor, $\Upsilon_0$ is the short-memory or viscoelastic coefficient, and $\mathbf{e}= \left(\nabla\mathbf{V}+\nabla\mathbf{V}^{T}\right)/2$ is the rate-of-strain tensor. The convected derivative used in the present model is
\begin{equation}\label{eq:convected_derivative}
    \frac{\delta\mathbf{e}}{\delta t}
    =\frac{\partial\mathbf{e}}{\partial t}
    +\mathbf{V}\cdot\nabla\mathbf{e}
    -\mathbf{e}\cdot\nabla\mathbf{V}
    \mathbin{{-}}(\nabla\mathbf{V})^{T}\cdot\mathbf{e}.
\end{equation}
When $\Upsilon_0=0$, Eq.~\eqref{eq:constitutive_dimensional} reduces to the Newtonian constitutive relation.
{The short-memory approximation requires the product of the material time $\Upsilon_0/\mu$ with the relevant deformation rates and frequencies to be small. Modes outside this range cannot be interpreted as physical predictions of the constitutive approximation.}

The computations below use dimensionless comparison families rather than a calibration to a particular polymer solution. A dimensional realization requires compatible thermal, surfactant and rheological coefficients.

\subsection{Governing equations}\label{subsec:governing_equations}

The fluid is assumed incompressible. The dimensional equations governing mass conservation, momentum balance, heat transport and bulk surfactant transport are
\begin{subequations}\label{eq:dimensional_governing_equations}
\begin{align}
    \nabla\cdot\mathbf{V} &=0, \label{eq:dim_continuity}\\
    \rho\left(\frac{\partial\mathbf{V}}{\partial t}+\mathbf{V}\cdot\nabla\mathbf{V}\right)
    &=\rho\mathbf{g}+\nabla\cdot\boldsymbol{\tau}, \label{eq:dim_momentum}\\
    \frac{\partial T}{\partial t}+\mathbf{V}\cdot\nabla T
    &=\alpha_T\nabla^2T, \label{eq:dim_energy}\\
    \frac{\partial C}{\partial t}+\mathbf{V}\cdot\nabla C
    &=D_b\nabla^2C. \label{eq:dim_concentration}
\end{align}
\end{subequations}
Here, $\alpha_T= {k_T}/(\rho c_p)$ is the thermal diffusivity, and $D_b$ is the diffusivity of the surfactant in the bulk liquid.

\subsection{Boundary conditions}\label{subsec:boundary_conditions_dimensional}

At the solid wall $z=0$, the no-slip and no-penetration conditions are imposed. The wall is impermeable to the surfactant and is maintained at the prescribed temperature $T_w$. Thus,
\begin{equation}\label{eq:wall_boundary_dimensional}
    u=0,\qquad w=0,\qquad \frac{\partial C}{\partial z}=0,\qquad T=T_w,
    \qquad z=0.
\end{equation}

At the free surface $z=h(x,t)$, the normal and tangential stress balances are
\begin{align}
    p_a+\mathbf{n}\cdot\boldsymbol{\tau}\cdot\mathbf{n}
    &=-\sigma\nabla\cdot\mathbf{n}, \label{eq:normal_stress_dimensional}\\
    \mathbf{n}\cdot\boldsymbol{\tau}\cdot\mathbf{t}
    &=\nabla\sigma\cdot\mathbf{t}, \label{eq:tangential_stress_dimensional}
\end{align}
where $p_a$ is the ambient gas pressure. The unit normal and tangent vectors are
\begin{equation*}\label{eq:normal_tangent_dimensional}
    \mathbf{n}=\frac{(-h_x,1)}{\sqrt{1+h_x^2}},
    \qquad
    \mathbf{t}=\frac{(1,h_x)}{\sqrt{1+h_x^2}}.
\end{equation*}
The kinematic condition is
\begin{equation}\label{eq:kinematic_dimensional}
    \frac{\partial h}{\partial t}+\mathbf{V}\cdot\nabla(h-z)=0,
    \qquad z=h(x,t),
\end{equation}
which states that a material point on the interface remains on the interface. Heat exchange with the passive gas is described by Newton's law of cooling,
\begin{equation}\label{eq:newton_cooling_dimensional}
    k_T\nabla T\cdot\mathbf{n}+k_g(T-T_\infty)=0,
    \qquad z=h(x,t),
\end{equation}
where $k_g$ is the heat-transfer coefficient between the liquid film and the surrounding gas.

\subsection{Surfactant transport and surface tension law}\label{subsec:surfactant_surface_law}

Since the surfactant is soluble, two concentrations are required: the bulk concentration $C(x,z,t)$ and the surface concentration $\Gamma(x,t)$. The interfacial transport equation is written as \cite{pereiraTransportEquationInterfacial2008,dalessioMarangoniInstabilitiesAssociated2020}
\begin{equation}\label{eq:surface_transport_dimensional}
    {\partial_t^{\circ}\Gamma}
    +\nabla_s\cdot(\Gamma\mathbf{V}_s) + \Gamma(\nabla_s \cdot \mathbf{n})(\mathbf{V} \cdot \mathbf{n})
    =D_s\nabla_s^2\Gamma+J_{bs},
    \qquad z=h(x,t),
\end{equation}
{Here $D_s$ is the surface diffusivity, $\mathbf{V}_s=(\mathbf{I}-\mathbf{nn})\cdot\mathbf{V}$ is the tangential interfacial velocity, and $\partial_t^{\circ}$ denotes differentiation following the normal interface velocity $(\mathbf{V}\cdot\mathbf{n})\mathbf{n}$. The operator $\nabla_s=(\mathbf{I}-\mathbf{nn})\cdot\nabla$ is the surface gradient operator.} The quantity $J_{bs}$ is the net adsorption-desorption flux from the bulk to the interface. We use the {Langmuir adsorption--desorption law}
\begin{equation}\label{eq:adsorption_desorption_dimensional}
    J_{bs}=k_1\left(1-\frac{\Gamma}{\Gamma_\infty}\right)C_s-k_2\Gamma,
    \qquad C_s=C(x,h,t),
\end{equation}
where $k_1$ and $k_2$ are the adsorption and desorption rate constants, respectively, and $\Gamma_\infty$ is the maximum packing concentration on the interface. The same interfacial flux is related to the bulk concentration gradient by Fick's law,
\begin{equation}\label{eq:fick_flux_dimensional}
    J_{bs}=-D_b\nabla C\cdot\mathbf{n},
    \qquad z=h(x,t).
\end{equation}

The surface tension is assumed to depend linearly on both the surface surfactant concentration and the temperature. Thus,
\begin{equation}\label{eq:surface_tension_dimensional}
    \sigma=\sigma_0-\sigma_\Gamma(\Gamma-\Gamma_e)-\sigma_T(T-T_\infty),
\end{equation}
where $\sigma_0=\sigma(\Gamma_e,T_\infty)$, $\Gamma_e$ is the equilibrium surface concentration, and
\begin{equation*}
    \sigma_\Gamma=-\left.\frac{\partial\sigma}{\partial\Gamma}\right|_{\Gamma=\Gamma_e},
    \qquad
    \sigma_T=-\left.\frac{\partial\sigma}{\partial T}\right|_{T=T_\infty}.
\end{equation*}
For the usual case $\sigma_\Gamma>0$ and $\sigma_T>0$, an increase in either surfactant concentration or temperature decreases the surface tension.

\subsection{Non-dimensionalization}\label{subsec:nondimensionalization}

We introduce a long-wave scaling using the mean film thickness $\hat h_N$ as the transverse length scale and the disturbance wavelength $\hat\lambda$ as the streamwise length scale. The aspect ratio is
 $\varepsilon={2\pi\hat h_N}/{\hat\lambda}\ll 1$.
{
We choose the Nusselt free-surface velocity of the no-slip flat film as the characteristic velocity,
\[
    \hat V=\frac{g\sin\beta\,\hat h_N^2}{2\nu},
    \qquad \nu=\frac{\mu}{\rho}.
\]
Together with the Reynolds and Froude numbers defined below, this choice gives $Re/Fr^2=2$, the no-slip normalization used for comparison with our earlier falling-film study \cite{mukhopadhyaySolutocapillaryInstabilitySlipping2026}. We retain $Re/Fr^2$ explicitly in the governing equations and set it to $2$ when evaluating the results.
}
The variables are non-dimensionalized as
\begin{align}
(x^*,z^*,h^*)
&=
\left(
\frac{2\pi x}{\hat{\lambda}},
\frac{z}{\hat{h}_N},
\frac{h}{\hat{h}_N}
\right),
\qquad
(u^*,w^*)
=
\left(
\frac{u}{\hat{V}},
\frac{w}{\varepsilon\hat{V}}
\right),
\notag\\[3pt]
t^*
&=
\frac{2\pi\hat{V}}{\hat{\lambda}}\,t,
\qquad
(p^*,T^*)
=
\left(
\frac{p}{\rho\hat{V}^2},
\frac{T-T_\infty}{\Delta T}
\right),
\notag\\[3pt]
(C^*,\Gamma^*)
&=
\left(
\frac{\hat h_N C}{\Gamma_\infty},
\frac{\Gamma}{\Gamma_\infty}
\right),
\notag\\[3pt]
(\tau_{xx}^*,\tau_{zz}^*)
&=
\frac{{\hat h_N}}{\varepsilon\mu\hat V}
(\tau_{xx},\tau_{zz}),
\qquad
(\tau_{xz}^*,\tau_{zx}^*)
=
\frac{\hat h_N}{\mu\hat V}
(\tau_{xz},\tau_{zx}),
\label{eq:nondimensional_variables}
\end{align}
where $\Delta T=T_w-T_\infty$. The superscript $*$ is dropped in the rest of the paper.

The dimensionless groups are
\begin{align}\label{eq:dimensionless_numbers}
    Re&=\frac{\hat V\hat h_N}{\nu},
    & Fr&=\frac{\hat V}{\sqrt{g\sin\beta\,\hat h_N}},
    & Pr&=\frac{\nu}{\alpha_T},
    \notag\\
    We&=\frac{\sigma_0}{\rho\hat V^2\hat h_N},
    & Pe_b&=\frac{\hat V\hat h_N}{D_b},
    & Pe_s&=\frac{\hat V\hat h_N}{D_s},
    & Bi&=\frac{k_g\hat h_N}{k_T},
    \notag\\
    M_1&=\frac{\sigma_\Gamma\Gamma_\infty}{\rho\hat V^2\hat h_N},
    & M_2&=\frac{\sigma_T\Delta T}{\rho\hat V^2\hat h_N},
    & k_s&=\frac{k_2\hat h_N}{\hat V},
    & \kappa&=\frac{k_1}{k_2\hat h_N}.
\end{align}
Here $Re$, $Fr$, $Pr$, $We$, $Pe_b$, $Pe_s$ and $Bi$ are the Reynolds, Froude, Prandtl, Weber, bulk P\'eclet, surface P\'eclet and Biot numbers, respectively. The parameters $M_1$ and $M_2$ are the solutal and thermal Marangoni numbers, $k_s$ is the dimensionless desorption rate, and $\kappa$ is the adsorption equilibrium parameter.

{
The definitions of $Re$ and $Fr$ give
\[
    \frac{Re}{Fr^2}
    =\frac{g\sin\beta\,\hat h_N^2}{\nu\hat V}
    =2,
    \qquad Fr^2=\frac{Re}{2}.
\]
Thus $Re/Fr^2$ is fixed by the velocity normalization and is not an additional independent control parameter. Its explicit appearance below keeps the origin of the gravity terms clear.

We also define the material viscoelastic parameter and the Kapitza number at this stage:
\begin{equation}\label{eq:upsilon_nu_Ka}
    \Upsilon_\nu
    =\frac{\Upsilon_0}{\rho}
    \left(\frac{g\sin\beta}{\nu^2}\right)^{2/3},
    \qquad
    Ka=\frac{\sigma_0}
    {\rho\nu^{4/3}(g\sin\beta)^{1/3}}.
\end{equation}
The Kapitza-number convention is the same as in Ref.~\cite{mukhopadhyaySolutocapillaryInstabilitySlipping2026}. The parameter $\Upsilon_\nu$ compares the material memory time $\Upsilon_0/\mu$ with the viscous--gravity time $\nu^{1/3}(g\sin\beta)^{-2/3}$. Both $\Upsilon_\nu$ and $Ka$ are independent of film thickness for a fixed liquid and inclination.

Their relations to the thickness-based groups are
\begin{equation}\label{eq:upsilon_nu_Ka_relation}
    \frac{\Upsilon_0}{\rho\hat h_N^2}
    =\frac{\Upsilon_\nu}
    {\left[Re\left(\dfrac{Re}{Fr^2}\right)\right]^{2/3}},
    \qquad
    We=\frac{Ka}{Re^{5/3}}
    \left(\frac{Re}{Fr^2}\right)^{1/3}.
\end{equation}
In particular, the elastic coefficient multiplying material derivatives is
\[
    Re\,\frac{\Upsilon_0}{\rho\hat h_N^2}
    =\frac{Re^{1/3}\Upsilon_\nu}
    {\left(Re/Fr^2\right)^{2/3}}.
\]
This coefficient is used throughout the dimensionless governing equations, the linear stability problem and the nonlinear reduction. No separate dimensionless viscoelastic parameter is needed. For the capillary number $Ca=\mu\hat V/\sigma_0$, the corresponding identity is
\[
    \frac{1}{Ca}=Re\,We
    =\frac{Ka}{Re^{2/3}}
    \left(\frac{Re}{Fr^2}\right)^{1/3}.
\]
}

With this scaling, the dimensionless governing equations are
\begin{subequations}\label{eq:nondimensional_governing_equations}
\begin{align}
    \frac{\partial u}{\partial x}+\frac{\partial w}{\partial z}&=0, \label{eq:nd_continuity}\\
    \varepsilon Re\left(\frac{\partial u}{\partial t}
    +u\frac{\partial u}{\partial x}
    +w\frac{\partial u}{\partial z}\right)
    &=\frac{Re}{Fr^2}
    +\left(\varepsilon^2\frac{\partial\tau_{xx}}{\partial x}
    +\frac{\partial\tau_{xz}}{\partial z}\right), \label{eq:nd_xmomentum}\\
    \varepsilon^2 Re\left(\frac{\partial w}{\partial t}
    +u\frac{\partial w}{\partial x}
    +w\frac{\partial w}{\partial z}\right)
    &=-\frac{Re}{Fr^2}\cot\beta
    +\varepsilon\left(\frac{\partial\tau_{zx}}{\partial x}
    +\frac{\partial\tau_{zz}}{\partial z}\right), \label{eq:nd_zmomentum}\\
    \varepsilon RePr\left(\frac{\partial T}{\partial t}
    +u\frac{\partial T}{\partial x}
    +w\frac{\partial T}{\partial z}\right)
    &=\varepsilon^2\frac{\partial^2T}{\partial x^2}
    +\frac{\partial^2T}{\partial z^2}, \label{eq:nd_energy}\\
    \varepsilon Pe_b\left(\frac{\partial C}{\partial t}
    +u\frac{\partial C}{\partial x}
    +w\frac{\partial C}{\partial z}\right)
    &=\varepsilon^2\frac{\partial^2C}{\partial x^2}
    +\frac{\partial^2C}{\partial z^2}. \label{eq:nd_bulk_surfactant}
\end{align}
\end{subequations}
The non-dimensional components of the stress tensor are
\begin{subequations}\label{eq:nd_stress_components}
\begin{align}
\tau_{xx}
&=-\frac{Re}{\varepsilon}p
+2\frac{\partial u}{\partial x}
-2\varepsilon {\frac{Re^{1/3}\Upsilon_\nu}{\left(Re/Fr^2\right)^{2/3}}}\Bigg[
\frac{\partial^2u}{\partial x\partial t}
+u\frac{\partial^2u}{\partial x^2}
+w\frac{\partial^2u}{\partial x\partial z}
\notag\\
&\hspace{2.9cm}
-2\left(\frac{\partial u}{\partial x}\right)^2
-\frac{\partial u}{\partial z}\left(
\frac{1}{\varepsilon^2}\frac{\partial u}{\partial z}
+\frac{\partial w}{\partial x}\right)
\Bigg], \\
\tau_{zz}
&=-\frac{Re}{\varepsilon}p
+2\frac{\partial w}{\partial z}
-2\varepsilon {\frac{Re^{1/3}\Upsilon_\nu}{\left(Re/Fr^2\right)^{2/3}}}\Bigg[
\frac{\partial^2w}{\partial z\partial t}
+w\frac{\partial^2w}{\partial z^2}
+u\frac{\partial^2w}{\partial x\partial z}
\notag\\
&\hspace{2.9cm}
-2\left(\frac{\partial w}{\partial z}\right)^2
-\frac{\partial w}{\partial x}\left(
\frac{\partial u}{\partial z}
+\varepsilon^2\frac{\partial w}{\partial x}\right)
\Bigg], \\
\tau_{xz}=\tau_{zx}
&=\left(\frac{\partial u}{\partial z}
+\varepsilon^2\frac{\partial w}{\partial x}\right)
-\varepsilon {\frac{Re^{1/3}\Upsilon_\nu}{\left(Re/Fr^2\right)^{2/3}}}\Bigg[
\frac{\partial^2u}{\partial z\partial t}
+\varepsilon^2\frac{\partial^2w}{\partial x\partial t}
+u\left(\varepsilon^2\frac{\partial^2w}{\partial x^2}
+\frac{\partial^2u}{\partial x\partial z}\right)
\notag\\
&\hspace{2.9cm}
+w\left(\varepsilon^2\frac{\partial^2w}{\partial x\partial z}
+\frac{\partial^2u}{\partial z^2}\right)
-2\frac{\partial u}{\partial z}\frac{\partial w}{\partial z}
-2\varepsilon^2\frac{\partial u}{\partial x}\frac{\partial w}{\partial x}
\Bigg].
\end{align}
\end{subequations}

The dimensionless wall conditions are
\begin{equation}\label{eq:nd_wall_boundary}
    u=0,
    \qquad w=0,
    \qquad \frac{\partial C}{\partial z}=0,
    \qquad T=1,
    \qquad z=0.
\end{equation}
At the free surface, all conditions below are evaluated at $z=h(x,t)$. The normal stress condition is
\begin{align}\label{eq:nd_normal_stress}
    p_a
    &+\frac{\varepsilon}{Re}
    \left[
    \varepsilon^2\tau_{xx}\left(\frac{\partial h}{\partial x}\right)^2
    -2\tau_{xz}\frac{\partial h}{\partial x}
    +\tau_{zz}
    \right]
    \left[1+\varepsilon^2\left(\frac{\partial h}{\partial x}\right)^2\right]^{-1}
    \notag\\
    &=\varepsilon^2\left[We-M_1(\Gamma-\Gamma_e)-M_2T\right]
    \frac{\partial^2h}{\partial x^2}
    \left[1+\varepsilon^2\left(\frac{\partial h}{\partial x}\right)^2\right]^{-3/2}.
\end{align}
The tangential stress condition is
\begin{align}\label{eq:nd_tangential_stress}
    &\left[
    \left(1-\varepsilon^2\left(\frac{\partial h}{\partial x}\right)^2\right)\tau_{xz}
    +\varepsilon^2(\tau_{zz}-\tau_{xx})\frac{\partial h}{\partial x}
    \right]
    \left[1+\varepsilon^2\left(\frac{\partial h}{\partial x}\right)^2\right]^{-1/2}
    \notag\\
    &\qquad =-\varepsilon Re\left[
    M_1\frac{\partial\Gamma}{\partial x}
    +M_2\left(\frac{\partial T}{\partial x}
    +\frac{\partial T}{\partial z}\frac{\partial h}{\partial x}\right)
    \right].
\end{align}
The kinematic condition is
\begin{equation}\label{eq:nd_kinematic}
    w=\frac{\partial h}{\partial t}+u\frac{\partial h}{\partial x}.
\end{equation}
The surface surfactant balance becomes

{
\begin{eqnarray}\label{eq:nd_surface_surfactant}
\varepsilon\left[
\frac{\partial (s_\varepsilon\Gamma)}{\partial t}
+\frac{\partial (s_\varepsilon\Gamma u_s)}{\partial x}
\right]
=\frac{\varepsilon^2}{Pe_s}\frac{\partial}{\partial x}
\left(\frac{1}{s_\varepsilon}\frac{\partial\Gamma}{\partial x}\right)
\nonumber\\
+s_\varepsilon k_s[\kappa(1-\Gamma)C-\Gamma].
\end{eqnarray}
Here $s_\varepsilon=\sqrt{1+\varepsilon^2 h_x^2}$ is the surface metric factor and $u_s=u(x,h(x,t),t)$ is the streamwise velocity evaluated at the interface. The bulk concentration in the exchange term is also evaluated at $z=h(x,t)$. This form accounts for the change of interfacial area as the free surface moves.
}
The bulk concentration flux condition at the interface is
\begin{equation}\label{eq:nd_bulk_flux}
 \frac{1}{Pe_b \sqrt{1+ \varepsilon^2\left(\frac{\partial h}{\partial x} \right)^2}}   \left( \varepsilon^2\frac{\partial h}{\partial x}\frac{\partial C}{\partial x}
    -\frac{\partial C}{\partial z} \right)
    =k_s\left[\kappa(1-\Gamma)C-\Gamma\right].
\end{equation}
Finally, Newton's cooling law gives
\begin{equation}\label{eq:nd_cooling_law}
    \frac{\partial T}{\partial z}
    -\varepsilon^2\frac{\partial T}{\partial x}\frac{\partial h}{\partial x}
    =-Bi\,T \sqrt{1+ \varepsilon^2\left(\frac{\partial h}{\partial x} \right)^2}.
\end{equation}

For brevity, write $F=Re/Fr^2$ and
$\Upsilon=\Upsilon_0/(\rho\hat h_N^2)=\Upsilon_\nu/(FRe)^{2/3}$.
The material time relative to the advective thickness scale is
\begin{equation}\label{eq:memory_validity}
 Wi=\frac{\Upsilon_0}{\mu}\frac{\hat V}{\hat h_N}
 =Re\,\Upsilon=\frac{\Upsilon_\nu Re^{1/3}}{F^{2/3}}.
\end{equation}
The short-memory approximation requires small products of this time scale with the relevant dimensionless deformation rates and frequencies. In particular, the flat-film maximum shear gives the necessary measure $F Wi\ll1$. This condition also restricts the interpretation of eigenvalues of the truncated constitutive model. An unrestricted search for the fastest eigenvalue can select a mode outside the assumed rheological regime.

\subsection{Base state}\label{subsec:base_state}

The basic state is a steady, unidirectional, flat-film solution. It is written as
\begin{equation}\label{eq:base_hydrodynamics}
    U_B(z)=\frac{Re}{Fr^2}\left(z-\frac{z^2}{2}\right),
    \qquad W_B(z)=0,
    \qquad h_B=1,
\end{equation}
with pressure
\begin{equation}\label{eq:base_pressure}
    P_B(z)=-\frac{\cot\beta}{Fr^2}(z-1),
\end{equation}
up to an arbitrary additive constant. The corresponding base temperature, bulk surfactant concentration and surface surfactant concentration are
\begin{equation}\label{eq:base_temperature_concentration}
    T_B(z)=1-\frac{Bi\,z}{1+Bi},
    \qquad
    C_B\equiv C_e=\frac{\Gamma_e}{\kappa(1-\Gamma_e)},
    \qquad
    \Gamma_B=\Gamma_e.
\end{equation}
The equilibrium bulk concentration follows from the adsorption-desorption balance
\begin{equation}
    \kappa(1-\Gamma_e)C_e-\Gamma_e=0.
\end{equation}
{The base velocity is independent of $\Upsilon_\nu$, which enters the linear perturbation dynamics and the nonlinear wave evolution through the elastic stress terms. With the chosen normalization, $U_B(1)=Re/(2Fr^2)=1$ and the base flow rate is $q_B=Re/(3Fr^2)=2/3$.}

Because the surface concentration is scaled by the maximum packing concentration, {$0\leq\Gamma_e<1$ for finite $\kappa>0$ and finite $C_e$}. If
\begin{equation}
    S_t=C_e+\Gamma_e
\end{equation}
denotes the total surfactant loading in the uniform state, then
\begin{equation}\label{eq:gamma_equilibrium}
    \Gamma_e=\frac{1}{2}\left[
    1+S_t+\frac{1}{\kappa}
    -\sqrt{(S_t-1)^2+\frac{2}{\kappa}(S_t+1)+\frac{1}{\kappa^2}}
    \right].
\end{equation}
{At fixed $S_t$, the limit $\kappa\rightarrow\infty$ gives $\Gamma_e\rightarrow\min(S_t,1)$ and $C_e\rightarrow\max(S_t-1,0)$. Thus the equilibrium inventory is entirely interfacial for $S_t<1$, whereas a non-zero bulk inventory remains for $S_t>1$. Strong adsorption alone does not establish the full dynamical insoluble-surfactant limit.}

\section{Linear stability analysis}\label{sec:linear}

In this section we study the temporal stability of the uniform Nusselt-type base state derived in Sec.~\ref{subsec:base_state}. The analysis is carried out in two steps. First, the governing equations are linearized about the base state and written as an Orr-Sommerfeld-type eigenvalue problem. Second, the long-wave limit is used to obtain an analytical expression for the critical Reynolds number. Disturbances of arbitrary wavelength are then treated numerically using a Chebyshev spectral collocation method.

\subsection{Normal-mode formulation}\label{subsec:normal_mode}

Before forming the eigenvalue problem, we introduce
\[
    X=\frac{x_\ell}{\varepsilon},
    \qquad
    \tau=\frac{t_\ell}{\varepsilon},
    \qquad
    W=\varepsilon w_\ell,
\]
where $(x_\ell,t_\ell,w_\ell)$ denote the long-wave variables
introduced in Sec.~\ref{subsec:nondimensionalization}.
The streamwise and transverse coordinates are therefore both
scaled with $\hat h_N$, and time is scaled with
$\hat h_N/\hat V$.
We subsequently write $(x,t,w)$ for $(X,\tau,W)$.
The wavenumber $k$ below is consequently based on $\hat h_N$.

Let a small disturbance be superposed on the base state as
\begin{align}
    u(x,z,t)&=U_B(z)+\widetilde{u}(x,z,t), &
    w(x,z,t)&=\widetilde{w}(x,z,t), \nonumber\\
    p(x,z,t)&=P_B(z)+\widetilde{p}(x,z,t), &
    T(x,z,t)&=T_B(z)+\widetilde{T}(x,z,t), \nonumber\\
    C(x,z,t)&=C_B+\widetilde{C}(x,z,t), &
    \Gamma(x,t)&=\Gamma_B+\widetilde{\Gamma}(x,t), \nonumber\\
    h(x,t)&=1+\widetilde{h}(x,t).
    \label{eq:linear_perturbations}
\end{align}
Here the base quantities $U_B$, $P_B$, $T_B$, $C_B$ and $\Gamma_B$ are given in Sec.~\ref{subsec:base_state}, and the tildes denote infinitesimal perturbations. To satisfy continuity automatically, we introduce the perturbation stream function $\widetilde{\psi}$ through
\begin{equation}\label{eq:streamfunction_definition}
    \widetilde{u}=\frac{\partial \widetilde{\psi}}{\partial z},
    \qquad
    \widetilde{w}=-\frac{\partial \widetilde{\psi}}{\partial x}.
\end{equation}
The pressure perturbation is eliminated by combining the linearized streamwise and cross-streamwise momentum equations. We then introduce normal modes of the form
\begin{equation}\label{eq:normal_modes}
    \left(\widetilde{\psi},\widetilde{C},\widetilde{T},\widetilde{\Gamma},\widetilde{h}\right)
    =\left(\psi(z),\mathcal{C}(z),\Theta(z),\gamma,\eta\right)
    \exp\left[ik(x-\alpha t)\right],
\end{equation}
where $k>0$ is the real streamwise wavenumber and $\alpha=\alpha_r+i\alpha_i$ is the complex wave speed. With the convention in Eq.~\eqref{eq:normal_modes}, the temporal growth rate is $k\alpha_i$. Thus, a disturbance is unstable when $\alpha_i>0$, stable when $\alpha_i<0$, and neutrally stable when $\alpha_i=0$.

\subsection{Orr-Sommerfeld eigenvalue problem}\label{subsec:os_problem}

After substituting Eq.~\eqref{eq:normal_modes} into the linearized governing equations, the amplitudes $\psi(z)$, $\mathcal{C}(z)$ and $\Theta(z)$ satisfy the following coupled eigenvalue problem in $0<z<1$:
\begin{equation}\label{eq:os_momentum}
    \left[1-ik{\frac{Re^{1/3}\Upsilon_\nu}{\left(Re/Fr^2\right)^{2/3}}}\left(U_B-\alpha\right)\right]
    \left(D^2-k^2\right)^2\psi
    =ikRe\left[\left(U_B-\alpha\right)\left(D^2-k^2\right)-D^2U_B\right]\psi,
\end{equation}
\begin{equation}\label{eq:os_concentration}
    \left[\left(D^2-k^2\right)-ikPe_b\left(U_B-\alpha\right)\right]\mathcal{C}=0,
\end{equation}
\begin{equation}\label{eq:os_temperature}
    \left(D^2-k^2\right)\Theta
    -ikRePr\left[\left(U_B-\alpha\right)\Theta-DT_B\psi\right]=0,
\end{equation}
where $D\equiv d/dz$. {The elastic prefactor is obtained directly from Eq.~\eqref{eq:upsilon_nu_Ka_relation}; $\Upsilon_\nu$ is held fixed when $Re$ is varied at fixed material properties and inclination.} Equation~\eqref{eq:os_momentum} is the Orr-Sommerfeld equation modified by the weakly viscoelastic contribution from Walters' liquid B$''$. Equations~\eqref{eq:os_concentration} and \eqref{eq:os_temperature} describe the linearized bulk surfactant and thermal fields.

At the wall $z=0$, the no-slip, no-penetration, no-flux and fixed-temperature conditions give
\begin{equation}\label{eq:os_wall_bc}
    \psi=0,\qquad D\psi=0,\qquad D\mathcal{C}=0,\qquad \Theta=0.
\end{equation}
At the free surface $z=1$, the linearized tangential stress, normal stress, kinematic, thermal, interfacial surfactant and bulk-surfactant flux conditions are
\begin{align}
    &(D^2+k^2)\psi
    +ikRe\left[M_1\gamma+M_2\left(\Theta+DT_B\eta\right)\right]
    +D^2U_B\eta \nonumber \\
    &\hspace{1cm}
    -ik{\frac{Re^{1/3}\Upsilon_\nu}{\left(Re/Fr^2\right)^{2/3}}}\left[\left(U_B-\alpha\right)(D^2+k^2)\psi-D^2U_B\psi\right]=0,
    \label{eq:os_tangential_bc}
\end{align}
\begin{align}
    &(D^2-3k^2)D\psi
    -ikRe\left(U_B-\alpha\right)D\psi
    -ik\left[\frac{Re}{Fr^2}\cot\beta+k^2\left(
    {\frac{Ka}{Re^{2/3}}\left(\frac{Re}{Fr^2}\right)^{1/3}}
    -{Re}\,M_2T_B\right)\right]\eta \nonumber \\
    &\hspace{1cm}
    -ik{\frac{Re^{1/3}\Upsilon_\nu}{\left(Re/Fr^2\right)^{2/3}}}\left[\left(U_B-\alpha\right)(D^2-3k^2)D\psi+D^2U_BD\psi\right]=0,
    \label{eq:os_normal_bc}
\end{align}
\begin{equation}\label{eq:os_kinematic_bc}
    \psi+\left(U_B-\alpha\right)\eta=0,
\end{equation}
\begin{equation}\label{eq:os_thermal_bc}
    D\Theta+Bi\left(\Theta+DT_B\eta\right)=0,
\end{equation}
\begin{equation}\label{eq:os_surface_surfactant_bc}
    k^2\gamma
    +ikPe_s\left[\left(U_B-\alpha\right)\gamma+\Gamma_BD\psi\right]
    +k_sPe_s\left[(1+\kappa C_B)\gamma-\kappa(1-\Gamma_B)\mathcal{C}\right]=0,
\end{equation}
\begin{equation}\label{eq:os_bulk_flux_bc}
    D\mathcal{C}=Pe_bk_s\left[(1+\kappa C_B)\gamma-\kappa(1-\Gamma_B)\mathcal{C}\right].
\end{equation}
All quantities in Eqs.~\eqref{eq:os_tangential_bc}-\eqref{eq:os_bulk_flux_bc} are evaluated at $z=1$. The eigenvalue $\alpha$ determines the temporal behaviour of each disturbance mode. {The most dangerous admissible mode is the one with the largest value of $k\alpha_i$ among branches that remain within the short-memory regime. Selecting the largest growth rate from the unrestricted spectrum can instead select a fast mode outside the constitutive model's range of validity.}

\subsection{Long-wave asymptotic solution}\label{subsec:long_wave_solution}

The falling-film instability is often controlled by long waves. We therefore consider the limit $k\ll 1$ and expand the eigenfunctions and eigenvalue as
\begin{equation}\label{eq:longwave_expansion}
    \left(\psi,\mathcal{C},\Theta,\gamma,\eta,\alpha\right)
    =\left(\psi_0,\mathcal{C}_0,\Theta_0,\gamma_0,\eta_0,\alpha_0\right)
    +k\left(\psi_1,\mathcal{C}_1,\Theta_1,\gamma_1,\eta_1,\alpha_1\right)+O(k^2).
\end{equation}
The eigenfunctions are determined successively by substituting Eq.~\eqref{eq:longwave_expansion} into Eqs.~\eqref{eq:os_momentum}-\eqref{eq:os_bulk_flux_bc}. Since the eigenvalue problem is homogeneous, the amplitude of the free-surface displacement may be fixed without loss of generality. The neutral condition is obtained by imposing $\alpha_i=0$ at the first non-trivial order in $k$.

{

For the surface-wave branch, choose $\eta_0=1$ and define
$d=\kappa(1-\Gamma_e)^2$. The leading fields are
\begin{align}\label{eq:leading_longwave_fields}
 \alpha_0&=F, &\psi_0&=\frac{Fz^2}{2}, &
 \mathcal C_0&=\frac{6\Gamma_e}{4+3d},\\
 \gamma_0&=\frac{6\Gamma_e d}{4+3d}, &
 \Theta_0&=\frac{Bi^2z}{(1+Bi)^2}.
\end{align}
The first nonzero temporal growth coefficient follows as
\begin{align}\label{eq:longwave_growth_coefficient}
 \lambda(k)&=-ik\alpha(k)=-iFk+\mathcal G k^2+O(k^3),\\
 \mathcal G&=\frac{2}{15}Re F^2-\frac{F\cot\beta}{3}
 +\frac{Re\,\Upsilon F^2}{3}
 -\frac{3Re M_1\Gamma_e d}{4+3d}
 +\frac{Re M_2Bi}{2(1+Bi)^2}.
\end{align}
The long-wave threshold is $\mathcal G=0$. The thermal contribution is destabilizing for the stated heating convention, whereas the solutal term opposes growth. For fixed material $\Upsilon_\nu$, the elastic coefficient depends on $Re$ through Eq.~\eqref{eq:memory_validity}.

The long-wave neutral stability condition can be expressed in
terms of $\Upsilon_\nu$, defined in
Eq.~\eqref{eq:upsilon_nu_Ka}, as
\begin{equation}\label{eq:neutral_condition}
    \frac{\cot\beta}{Re_c}
    -\frac{2}{5}\left(\frac{Re}{Fr^2}\right)
    -\frac{\left(\frac{Re}{Fr^2}\right)^{1/3}\Upsilon_\nu}{Re_c^{2/3}}
    =f(Bi,\Gamma_e;M_1,M_2,\kappa),
\end{equation}
where
\[
    f(Bi,\Gamma_e;M_1,M_2,\kappa)
    =\left(\frac{Re}{Fr^2}\right)^{-1}
    \left[
    \frac{3M_2Bi}{2(1+Bi)^2}
    -\frac{9M_1\Gamma_e(1-\Gamma_e)^2\kappa}
    {4+3(1-\Gamma_e)^2\kappa}
    \right].
\]
The two terms in $f$ represent the thermal and solutal
Marangoni contributions, respectively.

}

\subsection{Critical Reynolds number}
\label{subsec:critical_reynolds}

{

Using Eq.~\eqref{eq:upsilon_nu_Ka_relation} at $Re=Re_c$,
the critical Reynolds number satisfies
\begin{equation}\label{eq:critical_reynolds_compact}
    Re_c=
    \frac{\cot\beta}{
    \frac{2}{5}\left(\frac{Re}{Fr^2}\right)
    +\dfrac{\left(\frac{Re}{Fr^2}\right)^{1/3}\Upsilon_\nu}{Re_c^{2/3}}
    +\left(\frac{Re}{Fr^2}\right)^{-1}
    \left[
    \dfrac{3M_2Bi}{2(1+Bi)^2}
    -\dfrac{9M_1\Gamma_e(1-\Gamma_e)^2\kappa}
    {4+3(1-\Gamma_e)^2\kappa}
    \right]
    }.
\end{equation}
At fixed $\Upsilon_\nu$, this is an implicit equation for
$Re_c$, because the thickness-based elastic coefficient in Eq.~\eqref{eq:upsilon_nu_Ka_relation} varies with $Re_c$. The ratio $Re/Fr^2$ remains fixed at $2$; at the threshold it is evaluated as $Re_c/Fr_c^2=2$, where $Fr_c$ denotes the Froude number at $Re=Re_c$.

For fixed $Re/Fr^2$, $Bi$, $\Gamma_e$, $M_1$, $M_2$
and $\kappa$, define
\[
    \mathcal{A}
    =\frac{2}{5}\left(\frac{Re}{Fr^2}\right)
    +\left(\frac{Re}{Fr^2}\right)^{-1}
    \left[
    \frac{3M_2Bi}{2(1+Bi)^2}
    -\frac{9M_1\Gamma_e(1-\Gamma_e)^2\kappa}
    {4+3(1-\Gamma_e)^2\kappa}
    \right].
\]
Then Eq.~\eqref{eq:critical_reynolds_compact} is equivalent to
\[
    \mathcal{A}Re_c
    +\left(\frac{Re}{Fr^2}\right)^{1/3}\Upsilon_\nu Re_c^{1/3}
    =\cot\beta.
\]
Thus, setting $y=Re_c^{1/3}$ gives the cubic equation
\[
    \mathcal{A}y^3+\left(\frac{Re}{Fr^2}\right)^{1/3}\Upsilon_\nu y
    -\cot\beta=0,
    \qquad Re_c=y^3.
\]
For $\mathcal{A}>0$, $\Upsilon_\nu\geq0$ and
$0<\beta<\pi/2$, this equation has a unique positive root.

In the Newtonian limit $\Upsilon_\nu=0$, the threshold reduces
to $Re_c=\cot\beta/\mathcal{A}$, provided $\mathcal{A}>0$.
In the clean isothermal limit $M_1=M_2=0$, the threshold satisfies
\begin{equation}\label{eq:critical_reynolds_clean_viscoelastic}
    \frac{2}{5}\left(\frac{Re}{Fr^2}\right)Re_c
    +\left(\frac{Re}{Fr^2}\right)^{1/3}\Upsilon_\nu Re_c^{1/3}
    =\cot\beta.
\end{equation}
For $\Upsilon_\nu=0$, this recovers
$Re_c=\tfrac{5}{2}(Re/Fr^2)^{-1}\cot\beta=\tfrac{5}{4}\cot\beta$.
}

\section{Nonlinear weighted-residual model}\label{sec:nonlinear}

The linear analysis of Sec.~\ref{sec:linear} describes the onset of instability and the growth of infinitesimal disturbances. To follow the subsequent nonlinear evolution, we derive a depth-averaged model in terms of the film thickness, flow rate, interfacial thermal variable, bulk surfactant variable and surface surfactant concentration. The derivation follows the long-wave weighted-residual philosophy introduced for falling films by Shkadov \cite{shkadovWaveFlowThin1967} and later refined by Ruyer-Quil and Manneville \cite{ruyerquilImprovedModelingFlows2000,ruyerquilFurtherAccuracyConvergence2002}. The model is constructed from the non-dimensional governing equations \eqref{eq:nondimensional_governing_equations} and boundary conditions \eqref{eq:nd_wall_boundary}-\eqref{eq:nd_cooling_law} obtained in Sec.~\ref{sec:formulation}.

\subsection{Long-wave reduction}\label{subsec:longwave_reduction}

We retain the terms required for a consistent long-wave description of inertia, viscosity, capillarity, Marangoni stresses, heat transfer, surfactant transport and weak viscoelasticity. The inertial contributions are kept up to $O(\varepsilon)$, while the streamwise viscous diffusion terms are retained up to $O(\varepsilon^2)$, following the standard weighted-residual ordering for falling films \cite{benneyLongWavesLiquid1966,kalliadasisFallingLiquidFilms2012}. The capillary contribution proportional to $\varepsilon^3 We$ is also retained, since it regularizes short waves and gives the small capillary ripples usually observed behind the main wave.

The pressure is obtained by integrating the cross-streamwise momentum equation \eqref{eq:nd_zmomentum} across the film and using the normal stress condition \eqref{eq:nd_normal_stress}. For the following pressure and momentum balances, we temporarily
return to the long-wave variables of
Sec.~\ref{subsec:nondimensionalization}, denoted here by
$(x,t,w)$, to display the asymptotic ordering explicitly.
This gives
\begin{equation}\label{eq:wrm_pressure_distribution}
    p=p_a+\frac{\cot\beta}{Fr^2}(h-z)
    -\frac{\varepsilon}{Re}\frac{\partial u}{\partial x}
    -\frac{\varepsilon}{Re}\left.\frac{\partial u}{\partial x}\right|_{z=h}
    -\varepsilon^2\left[We-M_1(\Gamma-\Gamma_e)-M_2T\right]\frac{\partial^2h}{\partial x^2}.
\end{equation}
Substitution of Eq.~\eqref{eq:wrm_pressure_distribution} into the streamwise momentum equation \eqref{eq:nd_xmomentum} yields the reduced momentum balance
 
\begin{eqnarray}\label{eq:wrm_reduced_momentum}
\varepsilon Re\left(\frac{\partial u}{\partial t}
+u\frac{\partial u}{\partial x}
+w\frac{\partial u}{\partial z}\right)
=\frac{Re}{Fr^2}
-\varepsilon\frac{Re}{Fr^2}\cot\beta\frac{\partial h}{\partial x}
+2\varepsilon^2\frac{\partial^2u}{\partial x^2}
+\varepsilon^2\frac{\partial}{\partial x}
\left[\left.\frac{\partial u}{\partial x}\right|_{z=h}\right]
+\frac{\partial^2u}{\partial z^2} {} \qquad  \nonumber\\
+\varepsilon^3 {\frac{Ka}{Re^{2/3}}\left(\frac{Re}{Fr^2}\right)^{1/3}}\frac{\partial^3h}{\partial x^3}\nonumber\\
+\varepsilon {\frac{Re^{1/3}\Upsilon_\nu}{\left(Re/Fr^2\right)^{2/3}}}\left[
4\frac{\partial u}{\partial z}\frac{\partial^2u}{\partial x\partial z}
-\frac{\partial}{\partial z}\left(
\frac{\partial^2u}{\partial t\partial z}
+u\frac{\partial^2u}{\partial x\partial z}
+w\frac{\partial^2u}{\partial z^2}
-2\frac{\partial u}{\partial z}\frac{\partial w}{\partial z}
\right)\right]
+\mathcal{O}(\varepsilon^3). {}\quad
\end{eqnarray}
 
After the long-wave truncation, we return to the thickness-based
variables used in the linear stability analysis:
\[
    X=\frac{x}{\varepsilon},
    \qquad
    \tau=\frac{t}{\varepsilon},
    \qquad
    W=\varepsilon w.
\]
Hence
\[
    \partial_x=\varepsilon^{-1}\partial_X,
    \qquad
    \partial_t=\varepsilon^{-1}\partial_\tau.
\]
The variables $h$, $u$, $q$, $T$, $C$ and $\Gamma$ retain
their original amplitude scales.
We subsequently omit the capital letters and write
$(x,t,w)$ for $(X,\tau,W)$.

This change of variables removes the explicit factors of
$\varepsilon$ from the retained equations. It does not change
the long-wave approximation used to derive them.
The distinguished ordering $We=O(\varepsilon^{-2})$ motivates
retention of the capillary term during the expansion; $We$
continues to denote the physical dimensionless group defined
in Eq.~\eqref{eq:dimensionless_numbers}.

The resulting reduced system is

\begin{subequations}\label{eq:wrm_reduced_system}
\begin{align}
 u_x+w_z&=0,\label{eq:wrm_continuity}\\
 Re(u_t+u u_x+w u_z)
 &=F-F\cot\beta\,h_x+2u_{xx}
   +\partial_x[u_x(x,h,t)]+u_{zz}\notag\\
 &\quad+\frac{Ka F^{1/3}}{Re^{2/3}}h_{xxx}
 +Wi\left[4u_z u_{xz}
 -\partial_z(u_{tz}+u u_{xz}+w u_{zz}-2u_z w_z)\right],
 \label{eq:wrm_momentum_reference}\\
 Pe_b(C_t+u C_x+w C_z)&=C_{xx}+C_{zz},\label{eq:wrm_bulk_concentration}\\
 RePr(T_t+u T_x+w T_z)&=T_{xx}+T_{zz}.\label{eq:wrm_temperature}
\end{align}
\end{subequations}

At the wall $z=0$,  equation ~\eqref{eq:nd_wall_boundary} gives
\begin{equation}\label{eq:wrm_wall_conditions}
    u=0, \qquad w=0, \qquad  \frac{\partial C}{\partial z}=0,
    \qquad T=1 \qquad .
\end{equation}
At the free surface $z=h(x,t)$, the reduced boundary conditions are
\begin{subequations}\label{eq:wrm_surface_conditions}

\begin{eqnarray}
    \frac{\partial u}{\partial z}
    =- \frac{\partial w}{\partial x}
    +4 \frac{\partial u}{\partial x}\frac{\partial h}{\partial x}
    +  {\frac{Re^{1/3}\Upsilon_\nu}{\left(Re/Fr^2\right)^{2/3}}}\left[
    2\left(\frac{\partial u}{\partial z}\right)^2\frac{\partial h}{\partial x}
    +\frac{\partial^2u}{\partial t\partial z}
    +u\frac{\partial^2u}{\partial x\partial z}
    +w\frac{\partial^2u}{\partial z^2}
    -2\frac{\partial u}{\partial z}\frac{\partial w}{\partial z}
    \right] \nonumber\\
     - Re\left[M_1\frac{\partial\Gamma}{\partial x}
    +M_2\left(\frac{\partial T}{\partial x}
    +\frac{\partial T}{\partial z}\frac{\partial h}{\partial x}\right)\right], \label{eq:wrm_tangential_condition}\\
    w =\frac{\partial h}{\partial t}+u\frac{\partial h}{\partial x}, \hspace{8cm} \label{eq:wrm_kinematic_condition}\\
    \frac{\partial\Gamma}{\partial t}
    +\frac{\partial}{\partial x}(u \Gamma)
    =\frac{ 1}{Pe_s}\frac{\partial^2\Gamma}{\partial x^2}
    +\frac{3}{ Pe_b}\frac{\chi}{h^2},
\hspace{4.5cm}\label{eq:wrm_conservative_surface_balance} \\
     \frac{\partial C}{\partial x}\frac{\partial h}{\partial x}
    -\frac{\partial C}{\partial z}
    =k_sPe_b\left[\kappa(1-\Gamma)C-\Gamma\right], \hspace{4cm}\label{eq:wrm_bulk_flux_condition}\\
    \frac{\partial T}{\partial z}
    - \frac{\partial T}{\partial x}\frac{\partial h}{\partial x}
    =-Bi\,T{\left[1+\frac{1}{2}\left(\frac{\partial h}{\partial x}\right)^2\right]}. \hspace{5cm} \label{eq:wrm_cooling_condition}
 \end{eqnarray}
\end{subequations}
The surface balance \eqref{eq:wrm_conservative_surface_balance} is written in the conservative form introduced in Sec.~\ref{subsec:conservative_closure} below.
{The surfactant closure uses the leading projected surface measure; surface-area corrections to surfactant storage and exchange are not included. Exact conservation below therefore refers to the projected reduced inventory, rather than to a complete second-order approximation of the curved-interface transport problem. In the cooling condition, the metric factor is expanded through the quadratic slope term required by the thermal equation.}

\subsection{Velocity, temperature and concentration ansatz}\label{subsec:wrm_ansatz}

We reduce the two-dimensional fields to depth-averaged variables by introducing the local coordinate
\begin{equation}
    \bar z=\frac{z}{h(x,t)}.
\end{equation}
The streamwise flow rate is defined as
\begin{equation}\label{eq:flow_rate_definition}
    q(x,t)=\int_0^h u(x,z,t)\,dz.
\end{equation}
The velocity field is projected onto polynomial basis functions satisfying the wall no-slip condition. Following the weighted-residual construction of Ruyer-Quil and Manneville \cite{ruyerquilImprovedModelingFlows2000,ruyerquilFurtherAccuracyConvergence2002}, {the expansion for homogeneous surface shear is}
\begin{equation}\label{eq:velocity_basis_general}
    u(x,z,t)=\sum_{i=0}^{K} a_i(x,t) {\widehat f_i}(\bar z),
    \qquad
    {\widehat f_i}(\bar z)=\bar z^{i+1}-\frac{i+1}{i+2}\bar z^{i+2}.
\end{equation}
{The present model uses $f_0=\widehat f_0$ and a separate zero-flux polynomial $f_1$ to represent the Marangoni shear. Thus $f_1$ below is not the $i=1$ member of the family $\widehat f_i$. The semi-parabolic approximation is}
\begin{equation}\label{eq:velocity_ansatz}
    u_0(x,z,t)=a_0(x,t)f_0(\bar z)+b_0(x,t)f_1(\bar z),
\end{equation}
where
\begin{equation}\label{eq:basis_functions_f0_f1}
    f_0(\bar z)=\bar z-\frac{1}{2}\bar z^2,
    \qquad
    f_1(\bar z)=2\bar z-3\bar z^2.
\end{equation}
The leading coefficient is fixed by the flow rate. With the gauge condition used below, one obtains
\begin{align}\label{eq:a0_b0_coefficients}
    a_0(x,t)
    &=\frac{q(x,t)}
    {h\int_0^1f_0(\bar z)\,d\bar z}
    =\frac{3q}{h},\\
    b_0(x,t)
    &=Re\frac{h}{4}
    \left(
    M_1\frac{\partial\Gamma}{\partial x}
    +M_2\frac{\partial\theta}{\partial x}
    \right).
\end{align}
where
\begin{equation}
    \theta(x,t)=T(x,h,t)
\end{equation}
is the interfacial temperature. The term proportional to $b_0$ represents the leading correction induced by tangential Marangoni stresses.

For the temperature field we use the linear profile
\begin{equation}\label{eq:temperature_ansatz}
    T(x,z,t)=1+(\theta-1)\bar z.
\end{equation}
This profile satisfies the wall temperature condition and defines the interfacial value $\theta$. The Newton cooling condition is then enforced in an averaged sense through the weighted energy equation.

For the soluble surfactant, the bulk concentration is approximated by
\begin{equation}\label{eq:bulk_concentration_ansatz}
    C(x,z,t)=c(x,t)+Pe_b k_s\frac{h}{2}\left[\kappa(1-\Gamma)c-\Gamma\right](1-\bar z^2),
\end{equation}
where
\begin{equation}\label{eq:surface_bulk_concentration_c}
    c(x,t)=C(x,h,t)
\end{equation}
is the bulk concentration evaluated at the free surface. {This profile satisfies the wall no-flux condition exactly and the leading normal-diffusion balance at the interface. It does not satisfy the slope-dependent condition \eqref{eq:wrm_bulk_flux_condition} pointwise to second order. Indeed, using $\chi$ as defined below,}
{
\[
    \left.h_xC_x-C_z\right|_{z=h}
    =\frac{3\chi}{h^2}+h_xc_x+\frac{3\chi}{h^2}h_x^2,
\]
whereas the kinetic term in Eq.~\eqref{eq:wrm_bulk_flux_condition}, combined with Eq.~\eqref{eq:chi_adsorption_relation}, is $3\chi/h^2$. The depth-integrated bulk balance must therefore be formed using the normal-flux condition before the unresolved profile is approximated. The profile then closes the bulk inventory and streamwise fluxes, without imposing an additional, incompatible pointwise surface condition.
}

The cross-streamwise velocity is obtained from continuity \eqref{eq:wrm_continuity} and the wall condition $w=0$:
\begin{equation}\label{eq:w_from_continuity}
    w(x,z,t)=-\int_0^z \frac{\partial u}{\partial x}(x,\zeta,t)\,d\zeta.
\end{equation}

\subsection{Bulk--surface closure and conservative fluxes}\label{subsec:conservative_closure}
Introduce the bulk excess concentration and the depth-integrated inventory,
\begin{equation}\label{eq:chi_definition}
 \chi=\int_0^h(C-c)\,\dd z,\qquad B=hc+\chi=\int_0^h C\,\dd z.
\end{equation}
The profile~\eqref{eq:bulk_concentration_ansatz} gives
\begin{equation}\label{eq:chi_adsorption_relation}
 \chi=\frac{Pe_bk_s h^2}{3}[\kappa(1-\Gamma)c-\Gamma].
\end{equation}
Thus the retained exchange rate is
\begin{equation}\label{eq:adsorption_source_chi}
 R=k_s[\kappa(1-\Gamma)c-\Gamma]=\frac{3\chi}{Pe_bh^2}.
\end{equation}
It enters the bulk and surface equations with opposite signs. This cancellation is necessary but must be supplemented by compatible streamwise fluxes.

To see the diffusion requirement, integrate the bulk equation before applying the polynomial closure. With an impermeable wall and the kinematic condition at $z=h$, the projected normal adsorption flux is
\begin{equation}\label{eq:projected_normal_flux}
 J=\frac{h_x C_x(h)-C_z(h)}{Pe_b},
 \qquad \int_0^h C_x\,\dd z=B_x-c h_x=\chi_x+h c_x.
\end{equation}
Consequently,
\begin{equation}\label{eq:integrated_bulk_identity}
 B_t+\partial_x\int_0^h uC\,\dd z
 =\frac{1}{Pe_b}\partial_x(\chi_x+h c_x)-J.
\end{equation}
The boundary contribution $h_xC_x(h)/Pe_b$ cannot be dropped while retaining streamwise diffusion on the moving liquid domain. For the quadratic concentration profile, it is
\begin{equation}\label{eq:bulk_diffusive_completion}
 \Delta_b=\frac{1}{Pe_b}\left(h_xc_x+\frac{3\chi h_x^2}{h^2}\right).
\end{equation}
It completes the older depth-integrated diffusion bracket into a streamwise divergence \cite{mukhopadhyayBulkSurfaceAudit2026}.

Define the interfacial Marangoni gradient by $\mathcal S=M_1\Gamma_x+M_2\theta_x$. Evaluation of the velocity profile gives
\begin{equation}\label{eq:surface_velocity_flux}
 u_s=\frac{3q}{2h}-\frac{Re\,h}{4}\mathcal S.
\end{equation}
The corresponding retained fluxes are
\begin{align}\label{eq:transport_fluxes}
 F_b&=cq+\frac{33q\chi}{40h}+\frac{3Re\,h\chi}{80}\mathcal S
       -\frac{\chi_x+h c_x}{Pe_b},\\
 F_s&=\frac{3q\Gamma}{2h}-\frac{Re\,h\Gamma}{4}\mathcal S
       -\frac{\Gamma_x}{Pe_s}.
\end{align}
The surface Marangoni term is a flux derivative. Its solutal contribution contains
$\partial_x(h\Gamma\Gamma_x)$, with coefficient one on
$\Gamma\Gamma_xh_x$ inside the derivative, rather than five in the earlier reduction \cite{mukhopadhyayConservationConsistentBoundary2026,mukhopadhyaySolutocapillaryInstabilitySlipping2026}.

Using the same reduced rate $R$ in both reservoirs gives
\begin{align}\label{eq:conservative_transport_pair}
 B_t+\partial_xF_b&=-R, & \Gamma_t+\partial_xF_s&=R,
\end{align}
and hence
\begin{equation}\label{eq:combined_transport}
 \partial_t(B+\Gamma)+\partial_x(F_b+F_s)=0.
\end{equation}
For periodic boundaries the projected inventory
\begin{equation}\label{eq:total_reduced_surfactant_mass}
 \mathcal M_s(t)=\int_0^L(B+\Gamma)\,\dd x
\end{equation}
is therefore constant. With nonperiodic boundaries its change is the net boundary flux. This property is algebraic and does not require adjustment of the mean concentration during time integration.

The quadratic concentration profile and the leading kinetic relation do not satisfy the complete finite-slope normal-flux condition pointwise to every geometric order. Exact conservation of~\eqref{eq:total_reduced_surfactant_mass} is a property of the stated projected model. The geometric inventory $\int_0^L[B+\Gamma\sqrt{1+h_x^2}]\,\dd x$ requires consistent surface-metric terms and a corresponding geometric extension of the closure.

The interfacial concentration can be recovered from $\chi$ as
\begin{equation}\label{eq:c_from_chi}
 c=\frac{\Gamma+3\chi/(Pe_b k_s h^2)}{\kappa(1-\Gamma)}.
\end{equation}
For computations it is preferable to evolve $B$ and use the equivalent recovery
\begin{equation}\label{eq:storage_recovery}
 a_s=\frac{Pe_bk_sh^2}{3},\qquad
 c=\frac{B+a_s\Gamma}{h+a_s\kappa(1-\Gamma)},\qquad \chi=B-hc.
\end{equation}
The denominator is positive for $h>0$, $0\leq\Gamma<1$ and positive kinetic parameters. Nonnegative reconstructed concentration is checked separately.

The thermal storage variable is
\begin{equation}\label{eq:phi_definition}
 \phi=\int_0^h[T(x,z,t)-\theta(x,t)]\,\dd z.
\end{equation}
The linear temperature profile gives
\begin{equation}\label{eq:phi_theta_relation}
 \phi=\frac12(1-\theta)h,\qquad \theta=1-\frac{2\phi}{h}.
\end{equation}
The computed state is $(h,q,\phi,B,\Gamma)$; $c$, $\chi$ and $\theta$ are recovered algebraically.
\subsection{Weighted-residual projection}\label{subsec:weighted_residual_projection}

The velocity is written as
\begin{equation}\label{eq:velocity_correction_decomposition}
    u=u_0+\widetilde{u},
\end{equation}
where $u_0$ is given by Eq.~\eqref{eq:velocity_ansatz} and
$\widetilde{u}$ denotes the retained first-order velocity
correction, with the explicit bookkeeping factor absorbed
into its definition.
Conservation of the flow rate gives the gauge condition
\begin{equation}\label{eq:gauge_condition}
    \int_0^h \widetilde{u}\,dz=0,
\end{equation}
which is standard in integral weighted-residual models \cite{samantaFallingFilmSlippery2011,samantaFallingFilmPorous2013}.

The residual of the reduced momentum equation is required to be orthogonal to a suitable weight function. In the Galerkin form this is written as
\begin{equation}\label{eq:weighted_residual_identity}
    \langle \mathcal{L}\widetilde{u},\mathcal{W}\rangle
    =\langle \widetilde{u},\mathcal{L}^{+}\mathcal{W}\rangle
    {+\left[\mathcal{W}\,\partial_z\widetilde{u}
    -(\partial_z\mathcal{W})\widetilde{u}\right]_{0}^{h}},
\end{equation}
{Here $\langle f,g\rangle=\int_0^h fg\,dz$, $\mathcal{L}=\partial_z^2$ and $\mathcal{L}^{+}=\partial_z^2$ is its formal adjoint. The boundary term generated by integration by parts must be retained because the surface condition on the velocity correction is inhomogeneous. The weight function is chosen as}
\begin{equation}\label{eq:momentum_weight_function}
    \mathcal{W}=f_0(\bar z).
\end{equation}
{For this weight, the gauge condition makes the volume term on the right-hand side of Eq.~\eqref{eq:weighted_residual_identity} vanish, while the boundary contribution is $\tfrac12\partial_z\widetilde{u}|_{z=h}$.}
The concentration equation is projected with weight $1$, while the thermal equation is projected with weight $z$. These choices give a five-equation model for $h$, $q$, $\phi$, $\chi$ and $\Gamma$. Similar polynomial profiles and integral projections have been used successfully in Newtonian falling-film models with thermal and surfactant effects \cite{oronLongscaleEvolutionThin1997,kalliadasisThermocapillaryInstability2003,trevelyanHeatedFallingFilms2007,dalessioMarangoniInstabilitiesAssociated2020}.

\subsection{Closed nonlinear system}\label{subsec:final_depth_averaged_model}
The hydrodynamic and thermal projections give the following equations in the thickness-based coordinates. Subscripts denote partial derivatives. The explicit long-wave bookkeeping parameter has been absorbed into the independent variables, while the retained asymptotic ordering remains that stated above.
\begin{equation}\label{model_mass}
    h_t+q_x=0.
\end{equation}
\begin{align}\label{model_q}
    q_t
    &+\partial_x\left[
    \frac{9}{7}\frac{q^2}{h}
    +{\frac{5}{12Re}\left(\frac{Re}{Fr^2}\right)\cot\beta}h^2
    +\frac{5}{4}\left(
    M_1\Gamma-2M_2\frac{\phi}{h}
    \right)
    \right]
    \notag\\
    &=
    \frac{1}{7}\frac{q}{h}q_x
    +{\frac{5Ka}{6Re^{5/3}}\left(\frac{Re}{Fr^2}\right)^{1/3}}
    h h_{xxx}
    +\frac{5}{2Re}
    \left(
    {\frac{1}{3}\left(\frac{Re}{Fr^2}\right)}h-\frac{q}{h^2}
    \right)
    \notag\\
    &\quad
    +\frac{1}{Re}\left[
    \frac{9}{2}q_{xx}
    -\frac{9}{2h}q_xh_x
    +4\frac{q}{h^2}h_x^2
    -6\frac{q}{h}h_{xx}
    \right]
    \notag\\
    &\quad
    +\frac{ReM_1}{16}\left[
    \frac{1}{3}h^2\Gamma_{tx}
    +\frac{15}{14}hq\Gamma_{xx}
    +\frac{19}{21}h q_x\Gamma_x
    +\frac{5}{7}q\Gamma_xh_x
    \right]
    \notag\\
    &\quad
    -\frac{ReM_2}{8}\left[
    \frac{1}{3}h\phi_{tx}
    -\frac{1}{3}\phi_t h_x
    +\frac{26}{21}\phi_xq_x
    -\frac{11}{7}\frac{\phi}{h}q_xh_x
    +\frac{1}{3}\phi q_{xx}
    \right.
    \notag\\
    &\qquad\left.
    +\frac{15}{14}q\phi_{xx}
    -\frac{10}{7}\frac{q}{h}\phi_xh_x
    +\frac{10}{7}\frac{q\phi}{h^2}h_x^2
    -\frac{15}{14}\frac{q\phi}{h}h_{xx}
    \right]
    \notag\\
    &\quad
    +{\frac{\Upsilon_\nu}
    {\left[Re\left(\dfrac{Re}{Fr^2}\right)\right]^{2/3}}}
    \left[
    \frac{5}{2h^2}q_t
    -15\frac{q^2}{h^4}h_x
    +15\frac{q}{h^3}q_x
    \right.
    \notag\\
    &\qquad
    +ReM_1\left(
    \frac{5}{4}\frac{q}{h^2}\Gamma_xh_x
    -\frac{5}{4h}q_x\Gamma_x
    -\frac{5}{8}\frac{q}{h}\Gamma_{xx}
    \right)
    \notag\\
    &\qquad\left.
    +2ReM_2\left(
    \frac{5}{2}\frac{\phi q}{h^4}h_x^2
    -\frac{5}{2}\frac{q}{h^3}\phi_xh_x
    -\frac{5}{4}\frac{\phi}{h^3}q_xh_x
    +\frac{5}{4h^2}q_x\phi_x
    -\frac{5}{8}\frac{\phi q}{h^3}h_{xx}
    +\frac{5}{8}\frac{q}{h^2}\phi_{xx}
    \right)
    \right].
\end{align}
The thermal balance is
\begin{align}\label{model_theta}
    \phi_t
    +\partial_x\left(
    \frac{27}{20}\frac{q\phi}{h}
    \right)
    &=
    \frac{7}{40}\frac{\phi}{h}q_x
    -\frac{3}{RePr}\left[
    Bi\left(-\frac{1}{2}+\frac{\phi}{h}\right)
    +\frac{\phi}{h^2}
    \right]
    \notag\\
    &\quad
    +\frac{1}{RePr}\left[
    \phi_{xx}
    -\frac{1}{h}\phi_xh_x
    -2\frac{\phi}{h}h_{xx}
    -\frac{3}{2}Bi
    \left(-\frac{1}{2}+\frac{\phi}{h}\right)h_x^2
    \right]
    \notag\\
    &\quad
    +\frac{3Re}{80}\left[
    M_1\left(
    h\phi\Gamma_{xx}
    +2h\phi_x\Gamma_x
    \right)
    \right.
    \notag\\
    &\qquad\left.
    -2M_2\left\{
    \phi\phi_{xx}
    +2\phi_x^2
    -4\frac{\phi}{h}\phi_xh_x
    -\frac{\phi^2}{h}h_{xx}
    +2\frac{\phi^2}{h^2}h_x^2
    \right\}
    \right].
\end{align}
The two surfactant equations are most compactly written using the fluxes in Eq.~\eqref{eq:transport_fluxes}:
\begin{align}\label{model_chi}
 B_t+\partial_xF_b&=-\frac{3\chi}{Pe_bh^2},\\
 \Gamma_t+\partial_xF_s&=\frac{3\chi}{Pe_bh^2}.\label{model_Gamma}
\end{align}
Equations~\eqref{model_mass}--\eqref{model_Gamma}, with Eqs.~\eqref{eq:storage_recovery} and~\eqref{eq:phi_theta_relation}, form the five-field model. Expanded bulk and surface balances are supplied in Appendix~\ref{app:expanded_transport}.

The mixed derivatives $\Gamma_{tx}$ and $\phi_{tx}$ in Eq.~\eqref{model_q} are obtained by differentiating their evolution right-hand sides. Collecting its remaining $q_t$ term yields the local coefficient
\begin{equation}\label{eq:time_coefficient}
 a_q(h)=1-\frac{5\Upsilon}{2h^2}
 =1-\frac{5\Upsilon_\nu}{2(FRe)^{2/3}h^2}.
\end{equation}
Explicit solution for $q_t$ requires $a_q\neq0$. This algebraic condition is distinct from the short-memory restriction in Eq.~\eqref{eq:memory_validity}.

\subsection{Uniform state and limiting cases}\label{subsec:limiting_cases}
The flat state is
\begin{align}\label{eq:wrm_base_state_consistency}
 h&=1, &q&=F/3, &\Gamma&=\Gamma_e,\\
 c&=C_e=\frac{\Gamma_e}{\kappa(1-\Gamma_e)}, &\chi&=0,
 &\phi&=\frac{Bi}{2(1+Bi)}.
\end{align}
It satisfies all five equations. Setting $\Upsilon_\nu=0$ gives the Newtonian stress response. Setting $M_1=M_2=0$ removes Marangoni feedback from the hydrodynamic problem; the scalar fields can remain passive. In particular, $M_2=0$ does not impose a uniform temperature. The sum of the scalar balances recovers Eq.~\eqref{eq:combined_transport} for every smooth periodic solution of the retained closure.

\section{Numerical formulation}\label{sec:numerics}
The linear problem is discretized on the interval $0\leq z\leq1$ by Chebyshev collocation \cite{gottliebNumericalAnalysisSpectral1977,trefethenSpectralMethodsMATLAB2000,schmidStabilityTransitionShear2001}. The mapping $\xi=2z-1$ gives the Gauss--Lobatto points
\begin{equation}
 \xi_j=-\cos(j\pi/N_z),\qquad z_j=(1+\xi_j)/2,\qquad 0\leq j\leq N_z.
\end{equation}
Differentiation with respect to $z$ is represented by $D_z=2D_\xi$. The momentum, concentration and thermal equations are imposed at the interior points, and the boundary conditions replace the appropriate rows. Including the surface concentration and displacement gives a generalized eigenvalue problem
\begin{equation}
 A\boldsymbol v=\alpha B\boldsymbol v,\qquad
 \boldsymbol v=(\psi_0,\ldots,\psi_{N_z},\mathcal C_0,\ldots,\mathcal C_{N_z},
 \Theta_0,\ldots,\Theta_{N_z},\gamma,\eta)^T.
\end{equation}
The matrices have order $3N_z+5$. A QZ decomposition determines the finite eigenvalues. With the convention $\exp[ik(x-\alpha t)]$, the temporal growth rate is $k\alpha_i$. The relevant surface-wave branch approaches $\alpha=F$ as $k\to0$. Its continuation, rather than an unrestricted maximization over the spectrum, is the appropriate comparison with the long-wave result. In a short-memory model, numerical convergence and constitutive validity are separate requirements.

For the nonlinear calculations, the streamwise domain is periodic with length $L=300$. The recorded discretization uses $N_x=1024$ points and a time increment $\Delta t=2\times10^{-4}$. The reference parameters are
\begin{align}
 Re&=15, & F&=2, &\beta&=10^\circ, &Ka&=6000,\\
 \Upsilon_\nu&=0.4, &M_1&=0.2, &M_2&=2, &Pr&=7,\\
 Bi&=0.1, &\Gamma_e&=0.5, &k_s&=1, &\kappa&=10,
 \qquad Pe_b=Pe_s=700.
\end{align}
The uniform reference state is given in Eq.~\eqref{eq:wrm_base_state_consistency}. For these parameters, $q_e=2/3$, $c_e=0.1$ and $\theta_e=10/11$. The figures follow the growth of a periodic disturbance and the subsequent development of its spatial profiles. Parameter comparisons are discussed through the displayed profiles and height histories. They are not used to infer wave speeds from differences in crest position.

The identity~\eqref{eq:combined_transport} is an analytical property of the conservative equations. A separate numerical test would evaluate the inventory from the stored fields throughout an integration. The profile figures presented below do not provide such an inventory history; conservation to a specified numerical tolerance is therefore not inferred from their appearance.

\section{Results and discussion}\label{sec:results}
\subsection{Linear instability and elastic memory}
The linear examples use $\beta=10^\circ$, $F=2$, $Pe_b=Pe_s=700$, $k_s=1$, $\kappa=10$, $\Gamma_e=0.5$, $Pr=7$, $Bi=0.1$ and $Ka=6000$. The Marangoni parameters and $\Upsilon_\nu$ are varied as indicated in the captions. Setting $M_1=M_2=0$ removes Marangoni feedback and isolates the hydrodynamic effect of elastic memory.

\begin{figure}[!htbp]
\begin{minipage}[t]{.48\linewidth}\centering
\includegraphics[width=\linewidth]{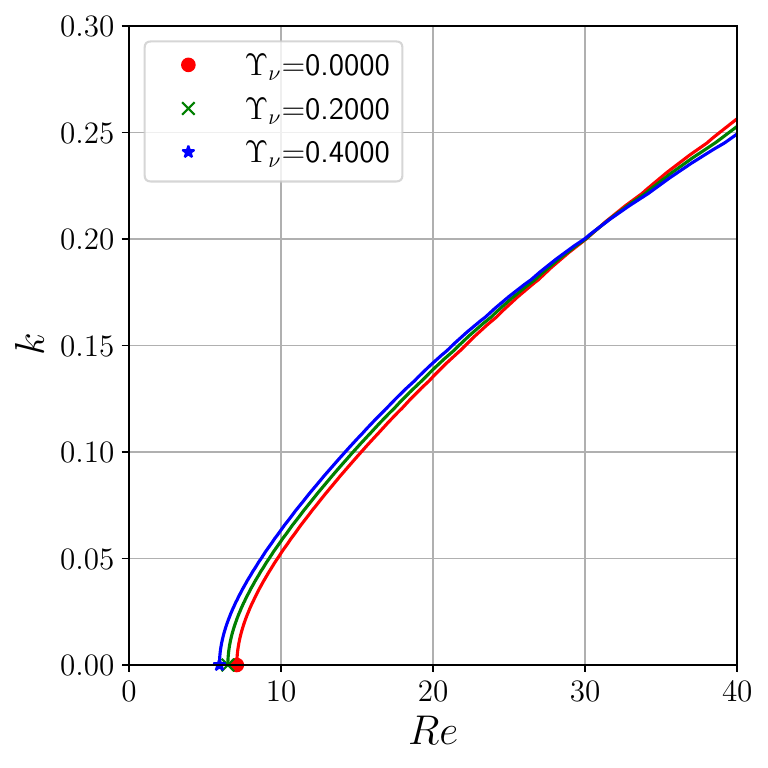}
\caption{Neutral curves for $M_1=M_2=0$ and $\Upsilon_\nu=0$, $0.2$ and $0.4$. The axes are Reynolds number and streamwise wavenumber. The markers on $k=0$ indicate the corresponding long-wave thresholds.}
\label{fig:clean_neutral}
\end{minipage}\hfill\begin{minipage}[t]{.48\linewidth}\centering
\includegraphics[width=\linewidth]{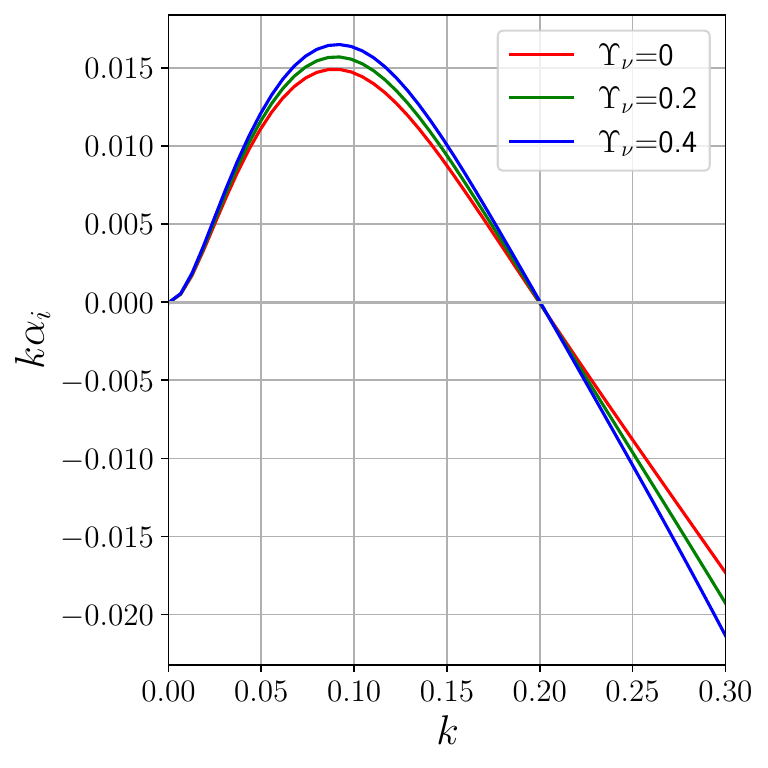}
\caption{Temporal growth rate $k\alpha_i$ for the clean-film problem at $Re=30$, with $M_1=M_2=0$ and $\Upsilon_\nu=0$, $0.2$ and $0.4$. Positive values denote growing disturbances.}
\label{fig:clean_growth}
\end{minipage}
\end{figure}

Figure~\ref{fig:clean_neutral} shows that the influence of memory depends on the Reynolds number. Increasing $\Upsilon_\nu$ lowers the long-wave onset, in agreement with the sign of the elastic contribution in Eq.~\eqref{eq:longwave_growth_coefficient}. At larger Reynolds numbers, the neutral curves cross. Thus the shift of the onset does not describe the entire neutral boundary. Beyond the crossing, increasing the elastic parameter reduces the limiting unstable wavenumber.

The growth-rate curves at $Re=30$ in Fig.~\ref{fig:clean_growth} help explain this dependence on wavelength. The larger elastic parameters increase the maximum amplification and have relatively little effect on the zero crossing near $k=0.2$. At higher wavenumbers, the same ordering produces stronger decay. Elastic memory therefore promotes the long-wave surface instability in this comparison while changing the damping of shorter disturbances in a different way.

\subsection{Thermal and solutal Marangoni stresses}
\begin{figure}[!htbp]
\centering
\includegraphics[width=.56\linewidth]{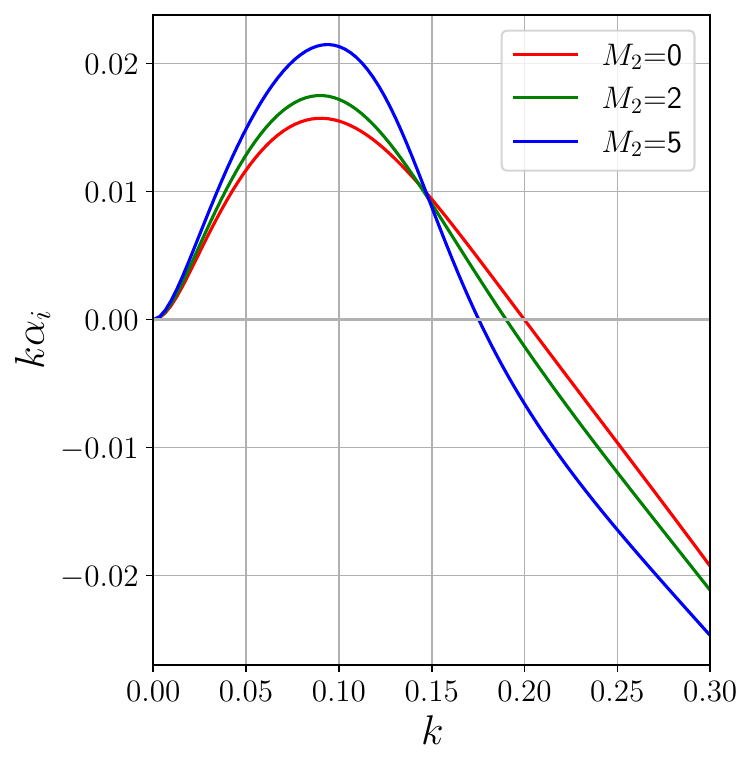}
\caption{Growth-rate curves at $Re=30$ for $M_2=0$, $2$ and $5$, with $M_1=0$ and $\Upsilon_\nu=0.2$. Increasing $M_2$ raises the peak amplification while reducing the high-wavenumber cutoff.}
\label{fig:thermal_growth}
\end{figure}

The thermal comparison is shown in Fig.~\ref{fig:thermal_growth}, where $M_1=0$. Raising $M_2$ increases the peak growth rate, as expected from the destabilizing thermal term in the long-wave expansion. The change in the cutoff is different: the positive-growth interval becomes shorter at its high-wavenumber end. Stronger amplification and a wider unstable band are therefore not equivalent descriptions of the thermal effect. Here thermal forcing strengthens the most amplified disturbances without extending instability to shorter waves.

\begin{figure}[!htbp]
\begin{minipage}[t]{.48\linewidth}\centering
\includegraphics[width=\linewidth]{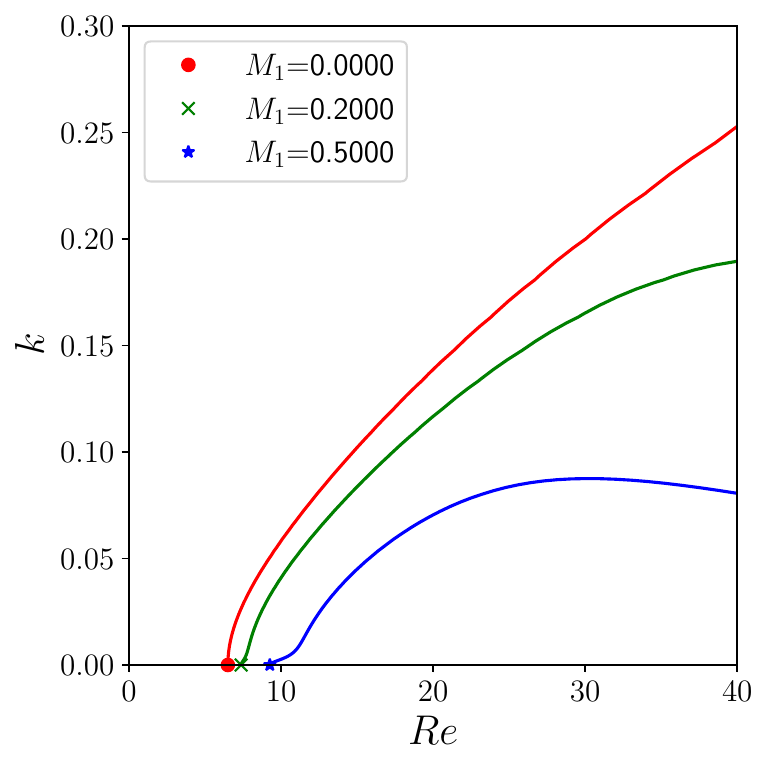}
\caption{Neutral curves for $M_1=0$, $0.2$ and $0.5$, with $M_2=0$ and $\Upsilon_\nu=0.2$. Stronger solutal Marangoni stress raises the onset and reduces the unstable region.}
\label{fig:solutal_neutral}
\end{minipage}\hfill\begin{minipage}[t]{.48\linewidth}\centering
\includegraphics[width=\linewidth]{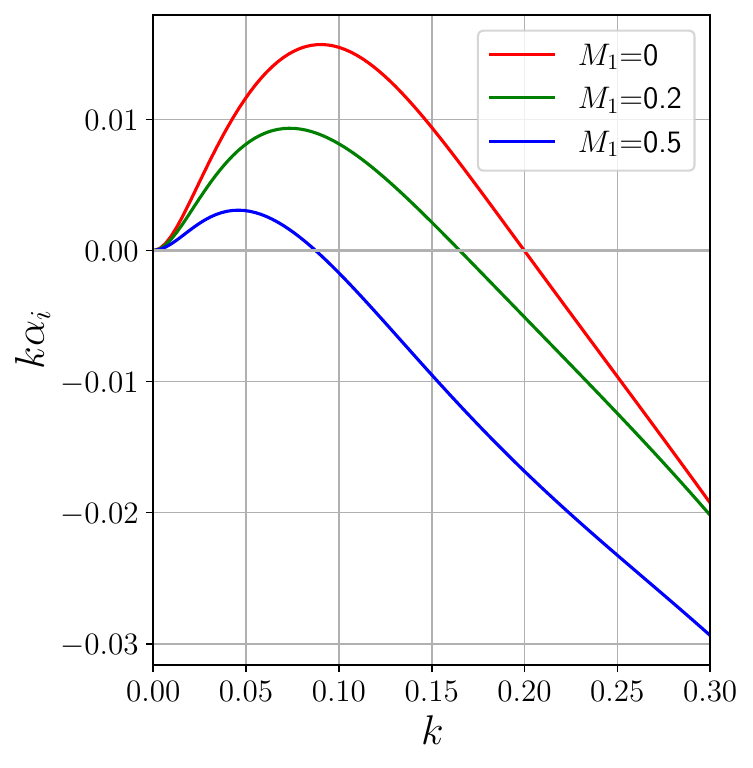}
\caption{Growth rates at $Re=30$ for $M_1=0$, $0.2$ and $0.5$, with $M_2=0$ and $\Upsilon_\nu=0.2$. The positive-growth band persists at the largest $M_1$, although its width and maximum growth rate are substantially reduced.}
\label{fig:solutal_growth}
\end{minipage}
\end{figure}

Figures~\ref{fig:solutal_neutral} and~\ref{fig:solutal_growth} show the corresponding solutal effect with $M_2=0$. Increasing $M_1$ shifts the onset to larger Reynolds numbers and reduces the range of unstable wavenumbers. The growth-rate maximum also decreases and moves towards longer waves. This response follows from the surface-concentration gradients generated by the disturbance: their Marangoni stress opposes the interfacial motion responsible for further amplification. For the largest value shown, a positive long-wave growth rate remains, so the film is not completely stabilized.

\begin{figure}[!htbp]
\centering
\includegraphics[width=.56\linewidth]{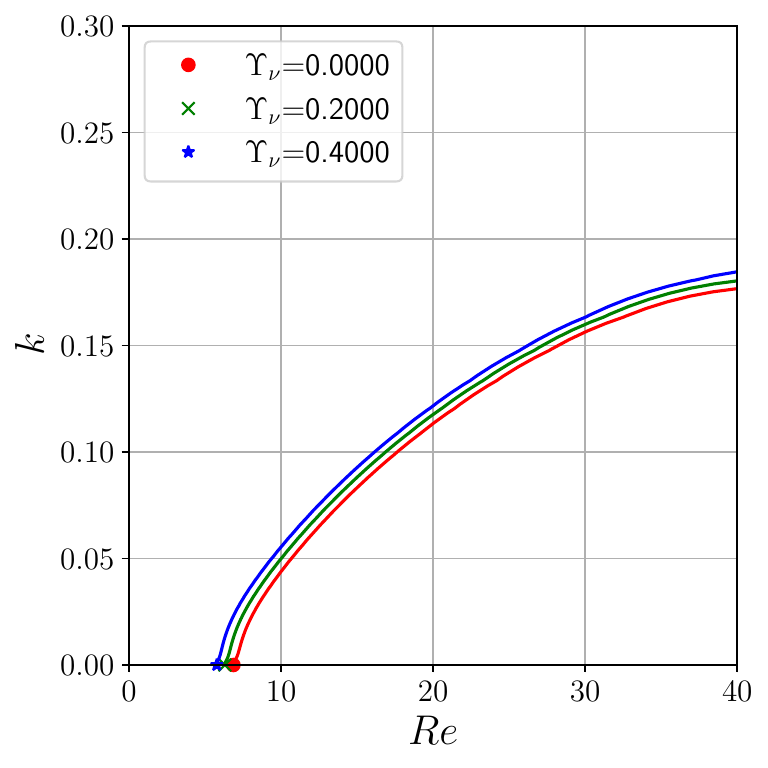}
\caption{Neutral curves with both Marangoni stresses present: $M_1=0.2$, $M_2=2$, and $\Upsilon_\nu=0$, $0.2$ and $0.4$.}
\label{fig:combined_neutral}
\end{figure}

When both stresses are present, the neutral boundary reflects their competition as well as the elastic contribution. Figure~\ref{fig:combined_neutral} gives the example $M_1=0.2$, $M_2=2$. Increasing $\Upsilon_\nu$ lowers the onset and raises the neutral wavenumber over the displayed Reynolds-number range. Unlike the clean-film example, no crossover is apparent in this range. This difference shows that the effect of elastic memory cannot be separated from the interfacial transport conditions by a single statement of stabilization or destabilization.

\subsection{Development of nonlinear waves}
\begin{figure}[!htbp]
\centering
\includegraphics[width=.98\linewidth]{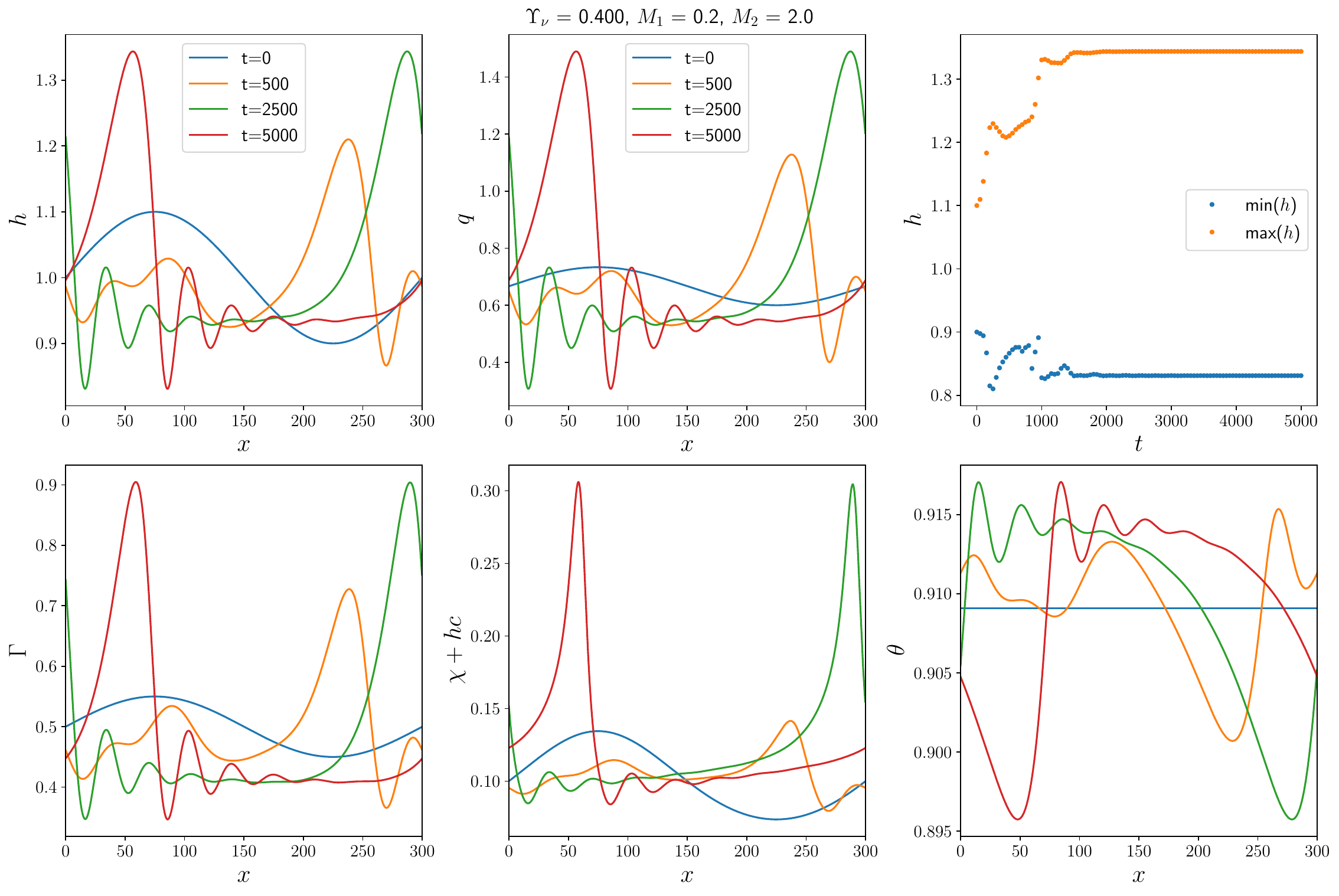}
\caption{Nonlinear development for $\Upsilon_\nu=0.4$, $M_1=0.2$ and $M_2=2$. The upper row shows film thickness, flow rate and the height-extrema histories; the lower row shows surface concentration, depth-integrated bulk inventory and interfacial temperature. The spatial profiles are shown at $t=0$, $500$, $2500$ and $5000$. The remaining parameters are stated in Sec.~\ref{sec:numerics}.}
\label{fig:nonlinear_development}
\end{figure}

Figure~\ref{fig:nonlinear_development} follows the reference case from the initially nearly uniform film to the later wave profiles. During the first part of the evolution, the surface disturbance grows and the maximum and minimum thickness separate. The growth then slows, and the extrema approach nearly constant levels. The spatial profiles show a pronounced hump and smaller capillary oscillations near the thinner part of the film. The displacement of the hump between snapshots is consistent with downstream propagation.

The flow-rate maximum accompanies the main surface hump, while the surface and bulk surfactant distributions become strongly nonuniform. Their profiles result from the combined action of advection, diffusion and adsorption--desorption exchange. The interfacial temperature also varies along the wave, although its relative variation is smaller than that of the concentration fields. The thermal field therefore continues to contribute to the local Marangoni stress after the height extrema have nearly saturated.

Nearly stationary extrema establish amplitude saturation over the displayed interval. They do not establish that every field translates with one constant speed. A permanent travelling wave would additionally require a common moving-frame profile, or an equivalent residual test of the travelling-wave equations. We therefore describe these states as developed nonlinear waves.

\subsection{Influence of the Marangoni stresses on the developed profiles}
\begin{figure}[!htbp]
\centering
\includegraphics[width=.98\linewidth]{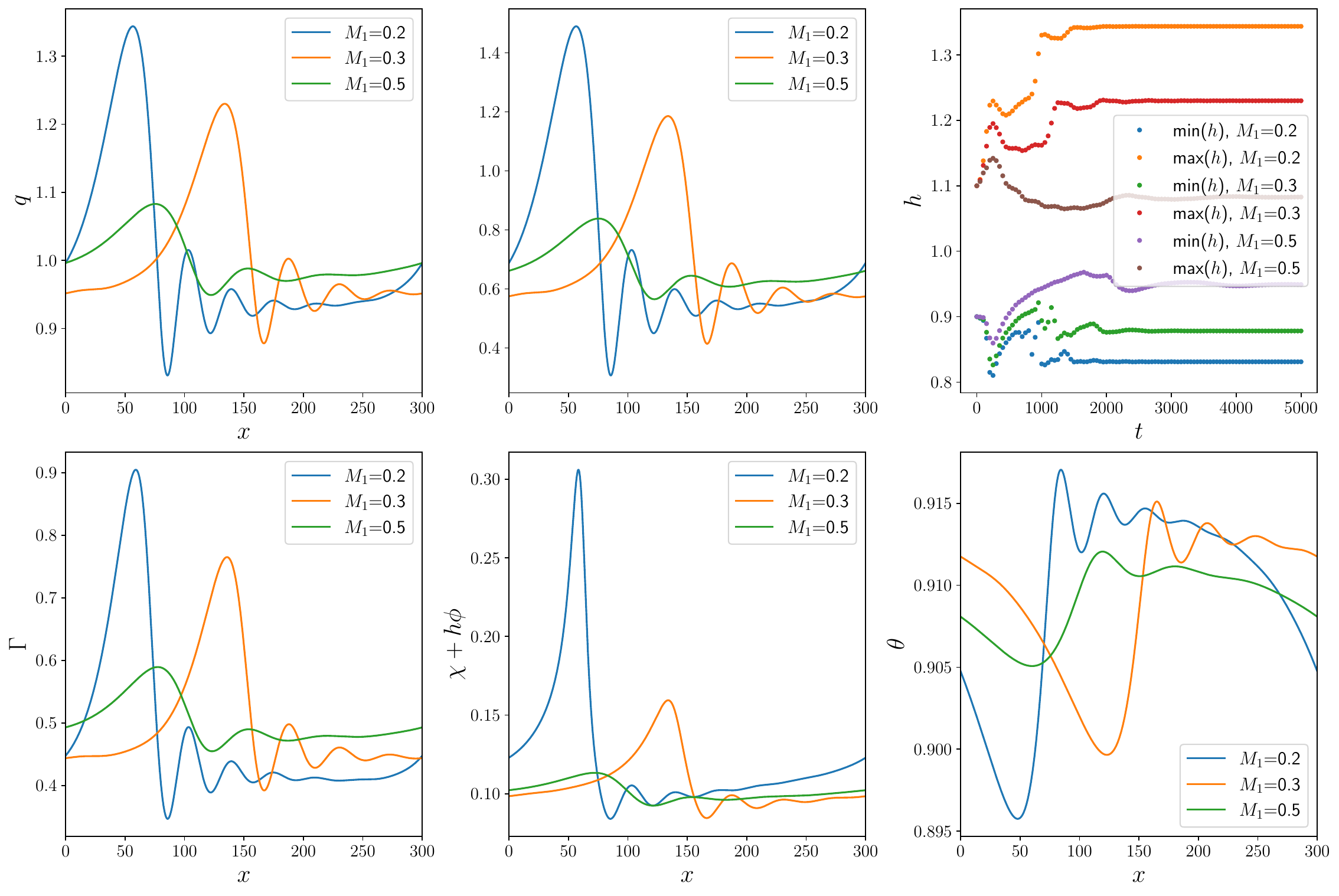}
\caption{Effect of $M_1=0.2$, $0.3$ and $0.5$ on the developed nonlinear profiles and height histories. Panels are arranged as in Fig.~\ref{fig:nonlinear_development}. In this original graphic, the upper-left ordinate denotes film thickness $h$ despite its printed $q$ label. The lower-middle ordinate represents $B=\chi+hc$; its printed concentration symbol differs from the notation used here. The profile comparison is interpreted at the displayed stage, without assigning an unrecorded common snapshot time.}
\label{fig:nonlinear_solutal}
\end{figure}

Figure~\ref{fig:nonlinear_solutal} compares $M_1=0.2$, $0.3$ and $0.5$. The reduction in wave amplitude as $M_1$ increases is evident in both the profiles and the height histories. The main hump becomes lower, the trough becomes shallower and the capillary ripples weaken. The peaks in flow rate and surfactant inventory also decrease. These changes are consistent with the linear suppression of the surface-wave branch, although linear growth rates alone do not determine the eventual wave amplitude.

The weaker thermal variations in the larger-$M_1$ cases accompany the smaller free-surface deformation. They should not be interpreted as a direct change of the imposed wall heating, since $M_1$ controls the solutal contribution to the surface stress. The comparison instead shows how changing one interfacial stress modifies the coupled hydrodynamic and transported fields.

\begin{figure}[!htbp]
\centering
\includegraphics[width=.98\linewidth]{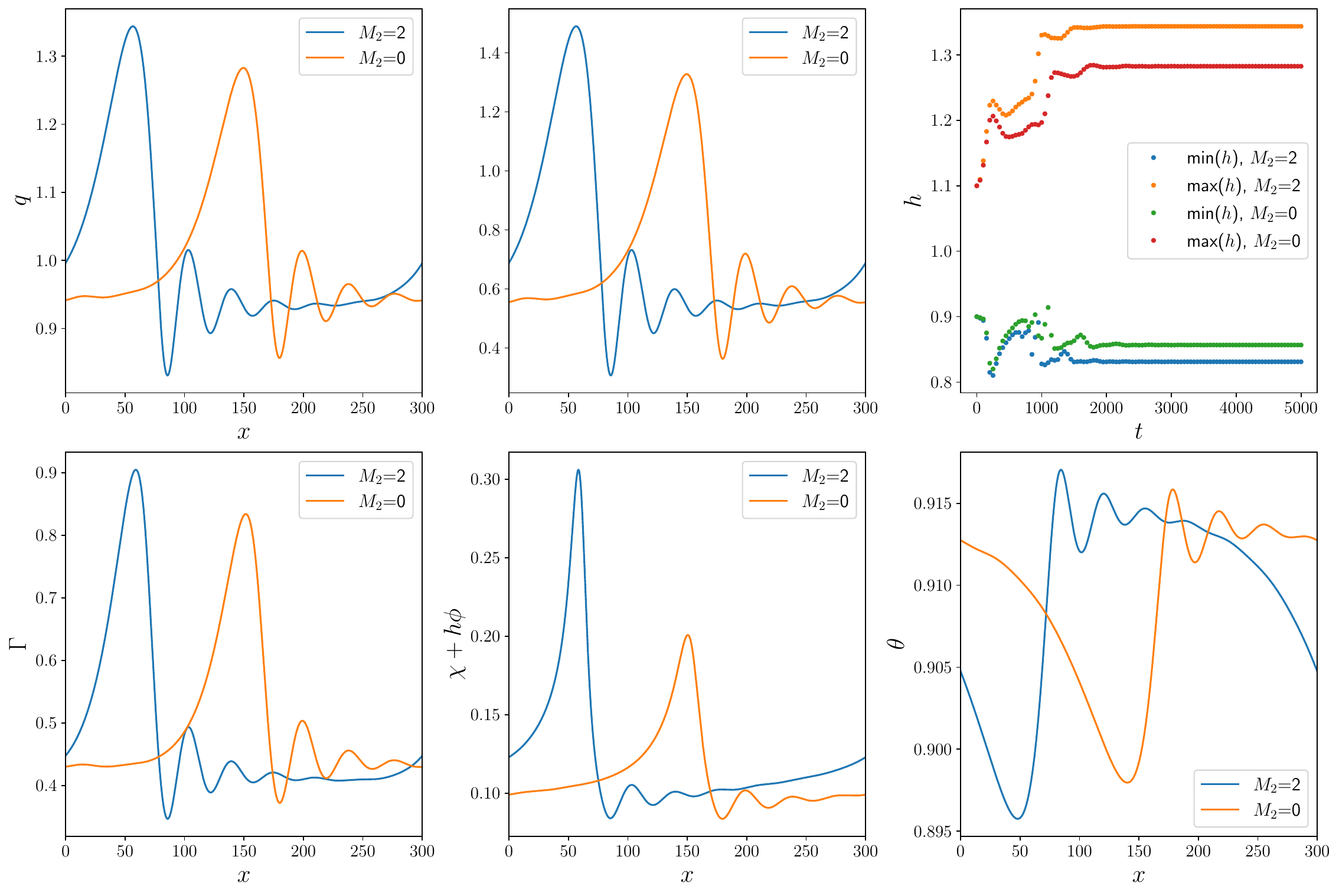}
\caption{Effect of thermocapillary feedback: $M_2=2$ and $M_2=0$. The panel arrangement is the same as in Fig.~\ref{fig:nonlinear_development}; the upper-left ordinate is $h$, and the lower-middle ordinate is the bulk inventory $B$, notwithstanding the original printed labels. Zero $M_2$ denotes zero thermal Marangoni feedback, with the temperature equation retained. No common profile time is assigned beyond the information displayed in the original figure.}
\label{fig:nonlinear_thermal}
\end{figure}

The cases $M_2=2$ and $M_2=0$ in Fig.~\ref{fig:nonlinear_thermal} separate the role of thermocapillary feedback. The nonzero-$M_2$ case has a larger height range and larger peaks in flow rate and surfactant inventory. Thermal Marangoni stress thus reinforces the nonlinear response in this example. This agrees with the increase in peak linear growth, but it does not imply that all thermal effects can be reduced to a change in the linear growth rate.

The case $M_2=0$ still has a nonuniform interfacial temperature. Setting this parameter to zero removes the temperature contribution to surface stress; it does not remove the thermal balance or impose an isothermal film. The distinction is necessary when comparing the temperature panels and interpreting the remaining surfactant-induced Marangoni stresses.

\subsection{Elastic contribution and the range of the model}
\begin{figure}[!htbp]
\centering
\includegraphics[width=.98\linewidth]{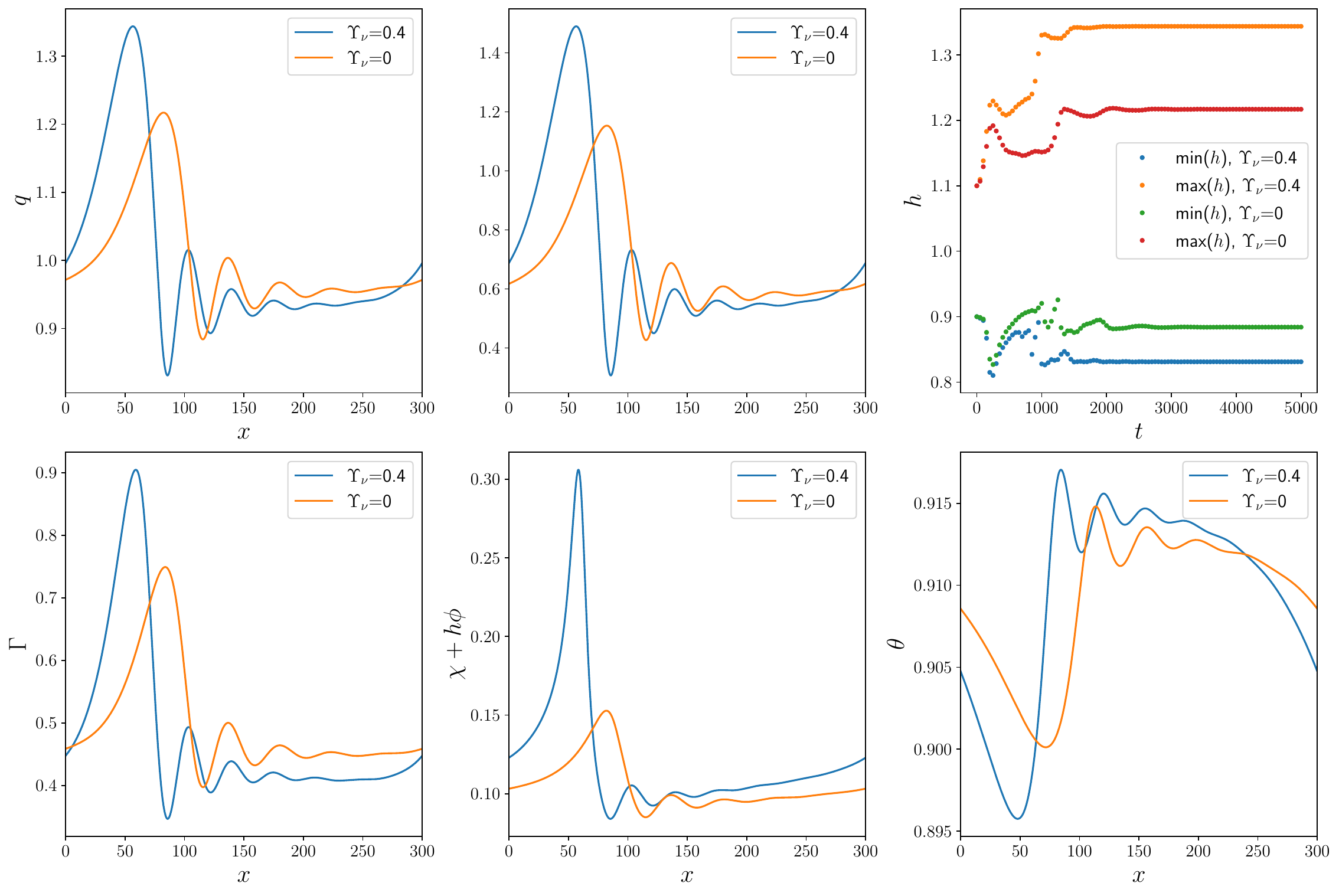}
\caption{Nonlinear comparison between $\Upsilon_\nu=0.4$ and the Newtonian limit $\Upsilon_\nu=0$, with $M_1=0.2$ and $M_2=2$. The upper-left profile is $h$ and the lower-middle profile is $B$, with the same original-label qualifications as in Fig.~\ref{fig:nonlinear_solutal}. The histories show amplitude development up to $t=5000$; the profile comparison is not used to infer a wave speed.}
\label{fig:nonlinear_memory}
\end{figure}

The Newtonian and viscoelastic examples are compared in Fig.~\ref{fig:nonlinear_memory}. The case $\Upsilon_\nu=0.4$ develops a larger surface deformation, stronger flow-rate and concentration peaks, and more pronounced capillary structure than the Newtonian case. The corresponding height histories also approach different late-time levels. The plotted crest positions differ, but these positions alone do not determine a difference in propagation speed.

The magnitude of the elastic parameter requires care. At $Re=15$, $F=2$ and $\Upsilon_\nu=0.4$, Eq.~\eqref{eq:memory_validity} gives $Wi\simeq0.62$ and $F Wi\simeq1.24$. The latter is not small. These nonlinear results are consequently formal predictions of the retained constitutive model, not an asymptotically controlled calculation for a weak-memory liquid at those parameters. The same restriction must be assessed along the linear branches at larger $\Upsilon_\nu$. It does not invalidate the algebraic Newtonian limit or the conservation identity, but it limits quantitative interpretation of the finite-parameter elastic trends.

There is a further distinction between transport conservation and the accuracy of a reduced solution. The flux construction ensures that adsorption and desorption redistribute surfactant without creating it. It does not establish the accuracy of the assumed depth profiles, the long-wave truncation or a particular discretization. Similarly, close agreement between wave shapes would not by itself test the combined inventory. These questions require separate checks when the model is used for quantitative prediction.

\section{Conclusions}
We have formulated a thermosolutal falling-film problem with soluble surfactant and a short-memory viscoelastic stress response. The linear analysis gives an Orr--Sommerfeld problem and a long-wave threshold in which thermal forcing, solutal Marangoni stress and elastic memory enter separately. The nonlinear reduction retains film thickness, flow rate, thermal storage, bulk surfactant inventory and surface concentration as independent fields. Compatible bulk and surface fluxes give a local balance for their combined projected inventory and hence exact analytical conservation on a closed periodic domain.

The numerical examples show different roles for the two Marangoni stresses. Increasing the solutal contribution raises the instability threshold, reduces growth rates and attenuates the developed waves. Thermal forcing increases peak linear growth and the nonlinear height range in the cases considered, although the high-wavenumber cutoff can decrease. Elastic memory shifts both the long-wave onset and the finite-wavelength response, and its effect depends on the thermal and surfactant conditions.

The nonlinear profiles develop a main hump with capillary ripples, together with substantial redistribution of surfactant between the bulk and interface. Their height histories approach nearly constant levels, which indicates amplitude saturation over the displayed interval. Establishing permanent travelling waves, numerical conservation to a prescribed tolerance and quantitative predictions for a particular liquid requires additional evidence beyond these profiles. The larger elastic cases should be read within the restrictions of the short-memory constitutive approximation. Within these limits, the formulation provides a conservative basis for studying the interaction between heat transfer, soluble-surfactant transport and elastic stress in falling films.

\appendix
\section{Expanded bulk and surface transport balances}\label{app:expanded_transport}
For comparison with the polynomial projection, the conservative bulk and surface equations may be expanded as follows. The bulk diffusion includes $h_xc_x$; the thermal Marangoni sign follows from $\theta=1-2\phi/h$.
\begin{align}\label{app:expanded_bulk}
    \partial_t(\chi+ch)
    &+\partial_x\left(
    \frac{33}{40}\frac{q\chi}{h}+cq
    \right)
    \notag\\
    &=
    \frac{1}{Pe_b}\left[
    \chi_{xx}
    {+h_xc_x}
    +h c_{xx}
    \right]
    -\frac{3}{Pe_b}\frac{\chi}{h^2}
    \notag\\
    &\quad
    -\frac{3Re}{80}\left[
    M_1\left(
    h\chi\Gamma_{xx}
    +h\chi_x\Gamma_x
    +\chi\Gamma_xh_x
    \right)
    \right.
    \notag\\
    &\qquad\left.
    \mathbin{{-}}2M_2\left\{
    \frac{\chi\phi}{h^2}h_x^2
    -\frac{1}{h}\left[
    h_x(\phi\chi_x+\chi\phi_x)
    +\chi\phi h_{xx}
    \right]
    +\chi_x\phi_x
    +\chi\phi_{xx}
    \right\}
    \right].
\end{align}
\begin{align}\label{app:expanded_surface}
    \Gamma_t
    +\frac{3}{2}\partial_x\left(
    \frac{q\Gamma}{h}
    \right)
    &=
    \frac{1}{Pe_s}\Gamma_{xx}
    +\frac{3}{Pe_b}\frac{\chi}{h^2}
    \notag\\
    &\quad
    +\frac{Re}{4}\left[
    M_1\left\{
    h\Gamma\Gamma_{xx}
    +h\Gamma_x^2
    +\Gamma\Gamma_xh_x
    \right\}
    \right.
    \notag\\
    &\qquad\left.
    -2M_2\left\{
    \Gamma_x\phi_x
    +\frac{\Gamma\phi}{h^2}h_x^2
    +\Gamma\phi_{xx}
    -\frac{1}{h}\left(
    \phi\Gamma_xh_x
    +\Gamma\phi_xh_x
    +\Gamma\phi h_{xx}
    \right)
    \right\}
    \right].
\end{align}

\section{Additional neutral-curve plot}\label{app:additional_neutral}
\begin{figure}[!htbp]
\centering
\includegraphics[width=.56\linewidth]{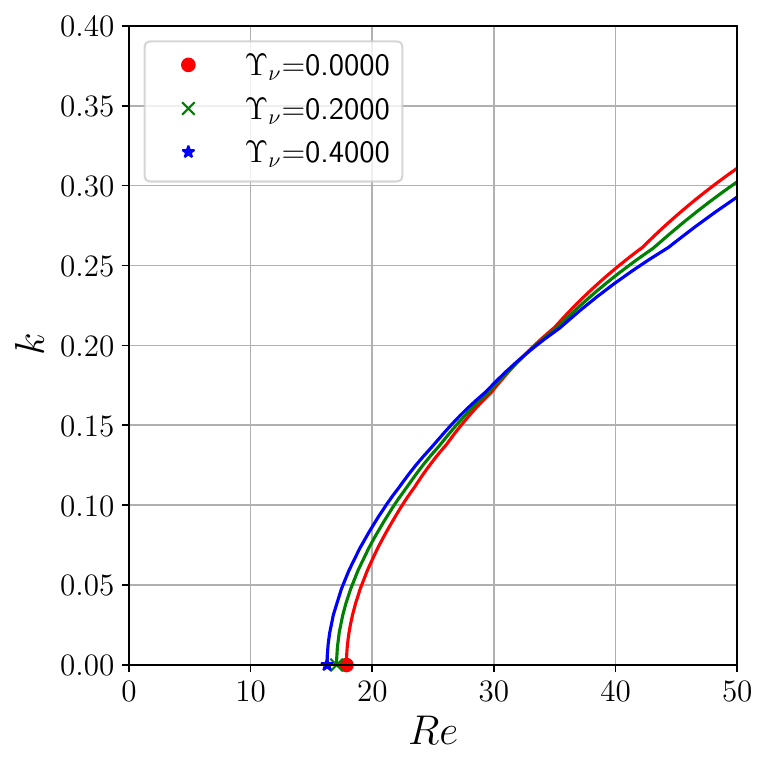}
\caption{Additional neutral curves, retained with their original legend: $\Upsilon_\nu=0$, $0.2$ and $0.4$. The plotted variables are $Re$ and $k$. No thermal-parameter ordering is inferred from this figure.}
\label{fig:additional_neutral}
\end{figure}

For completeness, Fig.~\ref{fig:additional_neutral} retains the additional neutral-curve calculation. Its legend identifies variation of $\Upsilon_\nu$. It is therefore not used as evidence for a variation of $M_2$ at fixed elasticity.

\FloatBarrier
\bibliographystyle{unsrtnat}
\bibliography{references}

@article{oronLongscaleEvolutionThin1997,
  author = {Oron, Alexander and Davis, Stephen H. and Bankoff, S. George},
  title = {Long-scale evolution of thin liquid films},
  journal = {Reviews of Modern Physics},
  volume = {69},
  number = {3},
  pages = {931--980},
  year = {1997},
  doi = {10.1103/RevModPhys.69.931}
}

@book{changComplexWaveDynamics2002,
  author = {Chang, Hsueh-Chia and Demekhin, Evgeny A.},
  title = {Complex Wave Dynamics on Thin Films},
  series = {Studies in Interface Science},
  volume = {14},
  publisher = {Elsevier},
  address = {Amsterdam},
  year = {2002}
}

@book{kalliadasisFallingLiquidFilms2012,
  author = {Kalliadasis, Serafim and Ruyer-Quil, Christian and Scheid, Benoit and Velarde, Manuel G.},
  title = {Falling Liquid Films},
  series = {Applied Mathematical Sciences},
  volume = {176},
  publisher = {Springer},
  address = {London},
  year = {2012},
  doi = {10.1007/978-1-84882-367-9}
}

@article{goussisSurfaceWaveThermocapillary1991,
  author = {Goussis, D. A. and Kelly, R. E.},
  title = {Surface wave and thermocapillary instabilities in a liquid film flow},
  journal = {Journal of Fluid Mechanics},
  volume = {223},
  pages = {25--45},
  year = {1991},
  doi = {10.1017/S0022112091001313}
}

@article{kalliadasisThermocapillaryInstability2003,
  author = {Kalliadasis, S. and Demekhin, E. A. and Ruyer-Quil, C. and Velarde, M. G.},
  title = {Thermocapillary instability and wave formation on a film falling down a uniformly heated plane},
  journal = {Journal of Fluid Mechanics},
  volume = {492},
  pages = {303--338},
  year = {2003},
  doi = {10.1017/S0022112003005809}
}

@article{ruyerquilThermocapillaryLongWaves2005,
  author = {Ruyer-Quil, C. and Scheid, B. and Kalliadasis, S. and Velarde, M. G. and Zeytounian, R. Kh.},
  title = {Thermocapillary long waves in a liquid film flow. {Part 1}. Low-dimensional formulation},
  journal = {Journal of Fluid Mechanics},
  volume = {538},
  pages = {199--222},
  year = {2005},
  doi = {10.1017/S0022112005005422}
}

@article{scheidThermocapillaryLongWaves2005,
  author = {Scheid, B. and Ruyer-Quil, C. and Kalliadasis, S. and Velarde, M. G. and Zeytounian, R. Kh.},
  title = {Thermocapillary long waves in a liquid film flow. {Part 2}. Linear stability and nonlinear waves},
  journal = {Journal of Fluid Mechanics},
  volume = {538},
  pages = {223--244},
  year = {2005},
  doi = {10.1017/S0022112005005446}
}

@article{trevelyanHeatedFallingFilms2007,
  author = {Trevelyan, P. M. J. and Scheid, B. and Ruyer-Quil, C. and Kalliadasis, S.},
  title = {Heated falling films},
  journal = {Journal of Fluid Mechanics},
  volume = {592},
  pages = {295--334},
  year = {2007},
  doi = {10.1017/S0022112007008476}
}

@article{pereiraTransportEquationInterfacial2008,
  author = {Pereira, A. and Kalliadasis, S.},
  title = {On the transport equation for an interfacial quantity},
  journal = {The European Physical Journal Applied Physics},
  volume = {44},
  number = {2},
  pages = {211--214},
  year = {2008},
  doi = {10.1051/epjap:2008103}
}

@article{karapetsasPrimaryInstabilityFalling2013,
  author = {Karapetsas, G. and Bontozoglou, V.},
  title = {The primary instability of falling films in the presence of soluble surfactants},
  journal = {Journal of Fluid Mechanics},
  volume = {729},
  pages = {123--150},
  year = {2013},
  doi = {10.1017/jfm.2013.291}
}

@article{pascalStabilityInclinedFlow2019,
  author = {Pascal, J. P. and D'Alessio, S. J. D. and Ellaban, E.},
  title = {Stability of inclined flow of a liquid film with soluble surfactants and variable mass density},
  journal = {Physical Review Fluids},
  volume = {4},
  pages = {054004},
  year = {2019},
  doi = {10.1103/PhysRevFluids.4.054004}
}

@article{dalessioMarangoniInstabilitiesAssociated2020,
  author = {D'Alessio, S. J. D. and Pascal, J. P. and Ellaban, E. and Ruyer-Quil, C.},
  title = {{Marangoni} instabilities associated with heated surfactant-laden falling films},
  journal = {Journal of Fluid Mechanics},
  volume = {887},
  pages = {A20},
  year = {2020},
  doi = {10.1017/jfm.2019.1058}
}

@article{dalessioLongwaveInstabilityThin2021,
  author = {D'Alessio, S. J. D. and Pascal, J. P.},
  title = {Long-wave instability in thin heated films doped with soluble surfactants},
  journal = {International Journal of Non-Linear Mechanics},
  volume = {136},
  pages = {103784},
  year = {2021},
  doi = {10.1016/j.ijnonlinmec.2021.103784}
}

@article{waltersMOTIONELASTICOVISCOUSLIQUID1960,
  author = {Walters, K.},
  title = {The motion of an elastico-viscous liquid contained between coaxial cylinders. {II}},
  journal = {The Quarterly Journal of Mechanics and Applied Mathematics},
  volume = {13},
  number = {4},
  pages = {444--461},
  year = {1960},
  doi = {10.1093/qjmam/13.4.444}
}

@article{beardElasticoviscousBoundarylayerFlows1964,
  author = {Beard, D. W. and Walters, K.},
  title = {Elastico-viscous boundary-layer flows. {I}. Two-dimensional flow near a stagnation point},
  journal = {Mathematical Proceedings of the Cambridge Philosophical Society},
  volume = {60},
  number = {3},
  pages = {667--674},
  year = {1964},
  doi = {10.1017/S0305004100038147}
}

@article{anderssonGravitydrivenFlowViscoelastic1999,
  author = {Andersson, H. I. and Dahl, E. N.},
  title = {Gravity-driven flow of a viscoelastic liquid film along a vertical wall},
  journal = {Journal of Physics D: Applied Physics},
  volume = {32},
  number = {14},
  pages = {1557--1562},
  year = {1999},
  doi = {10.1088/0022-3727/32/14/301}
}

@article{umaDynamicsThinViscoelastic2006,
  author = {Uma, B. and Usha, R.},
  title = {Dynamics of a thin viscoelastic film on an inclined plane},
  journal = {International Journal of Engineering Science},
  volume = {44},
  number = {20},
  pages = {1449--1481},
  year = {2006},
  doi = {10.1016/j.ijengsci.2006.08.014}
}

@article{mukhopadhyayWavesInstabilitiesViscoelastic2020,
  author = {Mukhopadhyay, S. and Mukhopadhyay, A.},
  title = {Waves and instabilities of viscoelastic fluid film flowing down an inclined wavy bottom},
  journal = {Physical Review E},
  volume = {102},
  pages = {023117},
  year = {2020},
  doi = {10.1103/PhysRevE.102.023117}
}

@article{amatousseTravelingWavesFalling2012,
  author = {Amatousse, N. and Ait Abderrahmane, H. and Mehidi, N.},
  title = {Traveling waves on a falling weakly viscoelastic fluid film},
  journal = {International Journal of Engineering Science},
  volume = {54},
  pages = {27--41},
  year = {2012},
  doi = {10.1016/j.ijengsci.2012.01.008}
}

@article{fuInstabilityWeaklyViscoelastic2018,
  author = {Fu, Qing-fei and Hu, Tao and Yang, Li-jun},
  title = {Instability of a weakly viscoelastic film flowing down a heated inclined plane},
  journal = {Physics of Fluids},
  volume = {30},
  pages = {084102},
  year = {2018},
  doi = {10.1063/1.5041494}
}

@article{sarmaThermosolutalMarangoni2021,
  author = {Sarma, R. and Mondal, P. K.},
  title = {Thermosolutal {Marangoni} instability in a viscoelastic liquid film: effect of heating from the free surface},
  journal = {Journal of Fluid Mechanics},
  volume = {909},
  pages = {A12},
  year = {2021},
  doi = {10.1017/jfm.2020.880}
}

@book{gottliebNumericalAnalysisSpectral1977,
  author = {Gottlieb, David and Orszag, Steven A.},
  title = {Numerical Analysis of Spectral Methods: Theory and Applications},
  publisher = {Society for Industrial and Applied Mathematics},
  address = {Philadelphia},
  year = {1977}
}

@book{trefethenSpectralMethodsMATLAB2000,
  author = {Trefethen, Lloyd N.},
  title = {Spectral Methods in {MATLAB}},
  publisher = {Society for Industrial and Applied Mathematics},
  address = {Philadelphia},
  year = {2000}
}

@book{schmidStabilityTransitionShear2001,
  author = {Schmid, Peter J. and Henningson, Dan S.},
  title = {Stability and Transition in Shear Flows},
  series = {Applied Mathematical Sciences},
  volume = {142},
  publisher = {Springer},
  address = {New York},
  year = {2001},
  doi = {10.1007/978-1-4613-0185-1}
}

@article{benneyLongWavesLiquid1966,
  author = {Benney, D. J.},
  title = {Long waves on liquid films},
  journal = {Journal of Mathematics and Physics},
  volume = {45},
  pages = {150--155},
  year = {1966},
  doi = {10.1002/sapm1966451150}
}

@article{shkadovWaveFlowThin1967,
  author = {Shkadov, V. Ya.},
  title = {Wave flow regimes of a thin layer of viscous fluid subject to gravity},
  journal = {Fluid Dynamics},
  volume = {2},
  number = {1},
  pages = {29--34},
  year = {1967},
  doi = {10.1007/BF01024797}
}

@article{ruyerquilImprovedModelingFlows2000,
  author = {Ruyer-Quil, C. and Manneville, P.},
  title = {Improved modeling of flows down inclined planes},
  journal = {The European Physical Journal B},
  volume = {15},
  pages = {357--369},
  year = {2000}
}

@article{ruyerquilFurtherAccuracyConvergence2002,
  author = {Ruyer-Quil, C. and Manneville, P.},
  title = {Further accuracy and convergence results on the modeling of flows down inclined planes by weighted-residual approximations},
  journal = {Physics of Fluids},
  volume = {14},
  number = {1},
  pages = {170--183},
  year = {2002},
  doi = {10.1063/1.1426103}
}

@article{samantaFallingFilmSlippery2011,
  author = {Samanta, A. and Ruyer-Quil, C. and Goyeau, B.},
  title = {A falling film down a slippery inclined plane},
  journal = {Journal of Fluid Mechanics},
  volume = {684},
  pages = {353--383},
  year = {2011},
  doi = {10.1017/jfm.2011.304}
}

@article{samantaFallingFilmPorous2013,
  author = {Samanta, A. and Goyeau, B. and Ruyer-Quil, C.},
  title = {A falling film on a porous medium},
  journal = {Journal of Fluid Mechanics},
  volume = {716},
  pages = {414--444},
  year = {2013},
  doi = {10.1017/jfm.2012.550}
}

@article{mukhopadhyayConservationConsistentBoundary2026,
  author = {Mukhopadhyay, S. and Millet, S. and Di Pierro, B. and Mukhopadhyay, A.},
  title = {Eliminating artificial surfactant-mass drift in nonlinear weighted-residual models of soluble-surfactant-laden falling films},
  journal = {Physics of Fluids},
  volume = {38},
  pages = {092118},
  year = {2026},
  doi = {10.1063/5.0344560}
}

@article{mukhopadhyaySolutocapillaryInstabilitySlipping2026,
  author = {Mukhopadhyay, S. and Millet, S. and Di Pierro, B. and Mukhopadhyay, A.},
  title = {Solutocapillary instability in slipping falling films},
  journal = {Physical Review Fluids},
  volume = {11},
  pages = {084004},
  year = {2026},
  doi = {10.1103/5xrv-cjwg}
}

@misc{mukhopadhyayBulkSurfaceAudit2026,
  author = {Mukhopadhyay, S. and Mukhopadhyay, A.},
  title = {A conservative bulk-surface closure for soluble-surfactant falling films: an analytical and numerical mass-balance audit},
  year = {2026},
  eprint = {2609.37869},
  archivePrefix = {arXiv},
  primaryClass = {physics.flu-dyn},
  url = {https://arxiv.org/abs/2609.37869},
  note = {arXiv:2609.37869}
}
\end{document}